\documentclass[12pt,a4paper]{article}
\pdfoutput=1
\usepackage{graphicx}
\usepackage{jheppub}
\usepackage{amsmath,amssymb,euscript,array,mathrsfs,epsfig}
\usepackage[small]{caption}
\usepackage{xcolor}

\def\ben
{\begin{equation}}
\def\een{\end{equation}}

    \let\L=\Lambda
   
 \let\W=\mu

\def\W={\cal W}
\def\L ={\cal L}

\def\be{\begin{equation}}
\def\ee{\end{equation}}
\def\ba{\begin{array}}
\def\ea{\end{array}}

\def\dalemb#1#2{{\vbox{\hrule height .#2pt
        \hbox{\vrule width.#2pt height#1pt \kern#1pt
                \vrule width.#2pt}
        \hrule height.#2pt}}}

\newcommand{\bea}{\begin{eqnarray}}
\newcommand{\eea}{\end{eqnarray}}

\title{Abelian Chern-Simons vortices at finite chemical potential}
\author{S. Prem Kumar, Stanislav Stratiev }
\affiliation{Department of Physics,\\
Swansea University,\\
Singleton Park, Swansea,\\
SA2 8PP, U.K.}
\emailAdd{s.p.kumar@swansea.ac.uk, stanislavstratiev@gmail.com}
\abstract{We examine vortices in Abelian Chern-Simons theory coupled to a relativistic scalar field  with a chemical potential for particle number or $U(1)$ charge. The Gauss constraint requires chemical potential for the local symmetry to be accompanied by a constant background charge density/magnetic field. Focussing attention on power law scalar potentials $\sim |\Phi|^{2s}$ which do not support vortex configurations in vacuum but do so at finite chemical potential,
we numerically study  classical vortex solutions for large winding number $|n|\gg 1$. The solutions depending on a single dimensionless parameter $\alpha$, behave as uniform incompressible droplets with radius $\sim \sqrt {\alpha |n|}$ , and energy scaling linearly with $|n|$, independent of  coupling constant. In all cases, the vortices transition from type I to type II at a critical value of the dimensionless parameter, $\alpha_c = \frac{s}{s-1}$, which we confirm with analytical arguments and numerical methods.
}

\begin{document}
\maketitle
\flushbottom
\section{Introduction}
Chern-Simons theory and the dynamics of vortices are both intimately connected to the Quantum Hall Effect \cite{Zhang:1988wy}. The Chern-Simons interaction  has the effect of attaching a magnetic flux to a charged particle, and endowing magnetic flux carrying vortices  with electric charge.  Magnetic flux attachment changes the  statistics of particles, so that fermions can be described as bosons with a Chern-Simons interaction. In particular, the effective theory of the fractional quantum hall state is a complex scalar interacting with an Abelian Chern-Simons gauge field \cite{Zhang:1988wy, Tong:2016kpv}.
Vortex solitons in the relativistic Abelian-Higgs model  in 2+1 dimensions in the presence  of a Chern-Simons action  (with or without a Maxwell term) have been extensively studied \cite{Paul:1986ix, Jackiw:1990aw, Hong:1990yh, Jackiw:1990pr, Dunne:1998qy, Horvathy:2008hd}.

In this note we are interested in vortex solitons appearing in  relativistic scalar field theory coupled to an Abelian Chern-Simons gauge field when a chemical potential for particle number is turned on. Our motivation is to investigate vortex configurations whose presence is triggered purely by finite density effects in Chern-Simons-matter theories. The Abelian Chern-Simons-scalar system offers the simplest  such setting. Eventually, we would like to understand finite density vortex solutions in $SU(N)$ and $U(N)$ Chern-Simons-scalar theories where  finite chemical potential results in condensation of gauge fields, potentially breaking rotational invariance \cite{Kumar:2018nkf}\footnote{An analogous situation in 3+1 dimensional Yang-Mills-Higgs system with $SU(2)\times U(1)$ gauge group was encountered in \cite{Gusynin:2003yu} where the finite density ground state breaks spatial isotropy due to condensation of vector fields.  Vortex solutions in the condensed phase were subsequently found in \cite{Gorbar:2005pi}.}.
A broader aim for exploring different aspects of finite density physics in Chern-Simons-matter theories is to understand the implications of the associated web \cite{Seiberg:2016gmd, Karch:2016sxi, Murugan:2016zal} of particle-vortex and Bose-Fermi dualities in 2+1 dimensions  
\cite{Giombi:2011kc, Aharony:2011jz, Aharony:2012nh, Aharony:2012ns, Jain:2013py, Jain:2013gza, Takimi:2013zca, Aharony:2015mjs, Geracie:2015drf}.

A chemical potential $\mu$ for a gauged $U(1)$ symmetry  can only be turned on provided a source term representing a classical (frozen out) uniform background charge density is simultaneously introduced. In Chern-Simons theory, such a source can either be viewed as a distribution of heavy charges or as a uniform magnetic field. We focus on the massless scalar field with purely power law interaction potential
\be 
U=\frac{g_{2s}}{s}\,|\Phi|^{2s}\,,\qquad s=2,3,\ldots\,, \label{powerlawu}
\ee
 with no symmetry breaking minima in vacuum, and solve the equations of motion numerically  with non-zero $\mu$.  Chern-Simons vortices with symmetry breaking potentials in vacuum have vanishing magnetic fields in the interior and on the outside, with the flux being supported at the edges. The finite $\mu$ vortex solutions are qualitatively different as the magnetic flux acquires support within the vortex interior, and depending on the sign of the quantized flux ($=2\pi n$), there are two qualitatively distinct types of solutions: (i) Those with negative flux, given our choice of conventions ($\mu>0$ and Chern-Simons level $k>0$), where the majority of the flux resides in the vortex interior, and (ii) positive flux solutions wherein most of the flux sits at the edge of the vortex. The qualitatively different behaviour of configurations with winding numbers $n<0$ and $n>0$ is expected due to the breaking of charge conjugation symmetry when $\mu\neq 0$.  Positive flux solutions are energetically disfavoured, or more precisely the grand potential for the $n$-vortex with $n>0$ is parametrically larger, as a function of $n$, than that for the $n<0$ vortex. 
 
The negative flux solutions are the most interesting. We find these numerically for a wide range of winding numbers, $1\leq |n| \lesssim 10^5$. Supported by simple analytical arguments, we confirm that for large $|n|$, negative flux vortices for power law potentials \eqref{powerlawu} with general $s$  exhibit linear scaling of the grand potential with $|n|$:
\be
{\cal E}(|n|\gg 1)\,=\,\frac{s-1}{2s}{k\mu} |n|\,.
\ee
For a specific value of the dimensionless coupling, the solutions are ``BPS" or marginally bound:
\be
\left.\frac{{\cal E}(|n|)}{|n|{\cal E}(1)}\right|_{\alpha = s/(s-1)}\,\to\, 1\,,\qquad \alpha\,\equiv\,
\frac{k}{4\pi}\left(\frac{g_{2s}}{\mu^{3-s}}\right)^{1/(s-1)}\,.
\ee
This critical value of $\alpha$ works surprisingly well even for low $n$ vortices. Below this value  individual vortices experience attractive interactions,  and repulsive interactions above it.  At the critical coupling, we find numerically that the vortex profiles closely (but not exactly) solve the first order Bogomolny'i type equation.
Finally, the radius of the $n$-vortex in all cases is given by
\be
\left.R_n\right|_{|n|\gg 1} \,=\,\sqrt{2\alpha |n|}\,\mu^{-1}\,,
\ee
implying  that the $n$-vortex behaves like a uniform incompressible droplet within which individual vortices are as closely packed as possible. The physical properties we have described  closely resemble the non-relativistic supersymmetric Chern-Simons theory introduced in \cite{Tong:2015xaa}. The scalings with $n$ are in line with the ``MIT bag" model for solitons with large winding number, advocated in \cite{Bolognesi:2005ty, Bolognesi:2005rk, Bolognesi:2007ez}.

The paper is organised as follows: In Sections 2 and 3, we review the  standard vortex equations of motion, but in the presence of chemical potential. We also point out some features of the finite density spectrum and qualitative aspects of vortex profiles.  In Section 4, we discuss the energy functional or the grand potential, and argue its expected scaling for large winding numbers. In Section 5, we present several results of the numerical analysis of the vortex equations of motion for different choices of potentials and parameters.
\section{The Abelian theory at finite chemical potential}
Our starting point is the Abelian Chern-Simons theory at level $k$ coupled to a relativistic scalar field in 2+1 dimensions. As is customary, we may regard this system as the  infrared limit of the Abelian-Higgs model with a Maxwell-Chern-Simons gauge field, since the Maxwell action is irrelevant compared to the Chern-Simons term. We want to consider the theory in the grand canonical ensemble with a chemical potential for the $U(1)$ charge. Turning on a chemical potential for a local symmetry is a subtle issue since the Gauss constraint requires the total charge in the system to vanish. This putative obstacle can be overcome by introducing  a uniform external classical charge density which can be viewed  as a distribution of  heavy charged species whose fluctuations are frozen out \cite{Kapusta:1981aa, Rosen:2010es}. In the presence of a Chern-Simons density this can also conveniently be viewed as a constant background magnetic field.

The $U(1)$ chemical potential is introduced as usual via a constant temporal background gauge field. Picking a $(-++)$ metric signature, the Lagrangian density for the system is, 
\bea
&&\mathcal{L} \,=\,\mathcal{L}_{\rm matter} \,+\,\mathcal{L}_{\rm CS}\, -\, J_0 A_0\,,
\\\nonumber\\\nonumber
&&\mathcal{L}_{\rm matter} \,=\, D_{\nu}\Phi^\dagger D^{\nu}\Phi + U(\Phi^\dagger\Phi) 
 \\\nonumber\\\nonumber
&&\mathcal{L}_{\rm CS} \,= \,\frac{k}{4 \pi} \epsilon^{\nu \lambda\sigma}
A_{\nu} \partial_{\lambda} A_{\sigma}\,.
\eea
The gauge-covariant derivatives on the complex scalar $\Phi$ include a background value $\mu$ for $A_0$, which is identified as a chemical potential for the $U(1)$ charge:
\be
D_{\nu}\, \equiv\, \partial_{\nu} -i A_{\nu}\,-\,i\mu\,\delta_{\nu,0}\,,
\ee
and $J_0$ is the background  classical charge density.  We assume  a  simple power law potential:
\be
U(\Phi^\dagger \Phi) \,=\, \frac{g_{2s}}{s} |\Phi|^{2s}\,, \qquad s \geq  2\,.
 \ee
 The cases of quartic ($s=2$) and sextic ($s=3$) potentials are special since they correspond to relevant and marginal interactions. Our main interest is in monotonic potentials with a global minimum at $\Phi=0$ (in the absence of a chemical potential), so that any stable vortex configurations only appear at finite density  i.e. they are driven by scalar condensation in the presence of the chemical potential. 
 
The quartic potential is of general interest because when $\mu$  vanishes, the theory flows to the 2+1 dimensional Wilson-Fisher fixed point coupled to a Chern-Simons gauge field. The critical scalar plays an important role in particle-vortex and the related web of Bose-Fermi dualities in 2+1 dimensions \cite{Aharony:2015mjs}. Semiclassical solutions are far removed from this critical point and only reliable when $\mu/g_4 \gg 1$.

 The Lagrangian density expanded to show the $\mu$-dependent terms is,
 \bea
 \mathcal{L} &&=\, \partial_{\nu} \Phi^{\dag}\partial^{\nu} \Phi \,+\,i A_{\nu} \left( \Phi^{\dag} \partial^{\nu} \Phi - \partial^{\nu}\Phi^{\dag} \Phi \right)\,+\,A_{\nu} A^{\nu} \left|\Phi \right|^2 +\frac{g_{2s}}{s}|\Phi|^{2s}- \mu^2 \left|\Phi \right|^2\nonumber \\\label{lagfull}
&&\,+\,i \mu\, \left(\Phi^{\dag} \partial^{0} \Phi -\partial^{0} \Phi^{\dag} \Phi  \right) \,-\, 2 \mu A_{0} \left|\Phi \right|^2 + \frac{k}{4 \pi} \epsilon^{\nu\lambda \sigma} A_{\nu} \partial_{\lambda} A_{\sigma}\, -\, J_0 A_0\,.
\eea
The background charge density $J_0$ is fixed by requiring that the expectation values of $A_0$ and the magnetic field vanish in the ground state:
\be
\langle A_0\rangle\,=\,0\,\implies\, J_0\,=\,-2\mu\langle|\Phi|\rangle^2\,,\qquad\qquad \langle|\Phi|\rangle\,=\,v\,=\,\left(\frac{\mu^2}{g_{2s}}\right)^{\frac{1}{2s-2}}\,.
\ee
The ground state conditions are also solved by a vanishing source $J_0=0$, and an  $A_0$ expectation value set by the chemical potential. This latter solution is equivalent to absorbing the chemical potential via a shift in the gauge field leaving the partition function unchanged. This possibility is excluded by imposing a vanishing $A_0$  at infinity as a boundary condition.

The classical source $J_0$ can also be interpreted as a background magnetic field. Defining
\be
B\,=\,\epsilon^{0ij}\partial_i A_j\,,
\ee 
and assuming a static configuration, the equation of motion for $A_0$ yields,
\be
J_0\,-\,\frac{k}{2\pi}\langle B\rangle\,+\,2\left(\mu\,+\,\langle A_0\rangle\right) v^2\,=\,0\,.
\ee
Therefore, even if the source $J_0$ were not  explicitly introduced, it is naturally induced through a non-zero background value for  $B$. We will treat this background value as distinct from the magnetic field carried by the vortex.

\subsection{Perturbative spectrum} The vacuum expectation value for $\Phi$ Higgses the gauge group and since Chern-Simons gauge fields do not propagate, the physical perturbative spectrum consists of two gapped excitations.  The dispersion relations can be found by expanding the gauge-fixed  action\footnote{We use an $R_\xi$ gauge fixing term of the form ${\cal L}_{\rm gf}\,=\,\left(\partial_\mu A^{\mu}\,-\,\xi \langle\Phi\rangle\delta\Phi^\dagger\,+\,\xi\delta\Phi\langle\Phi\rangle^\dagger\right)^2/(2\xi)$} to  quadratic order in fluctuations and identifying the gauge-invariant zeros of the fluctuation determinant. The dispersion relations for the two physical modes can be expressed in terms of dimensionless frequency and  momentum  and a single dimensionless coupling (assuming $k>0$, $\mu>0$)
\be
\alpha\equiv\, \frac{k\,\mu}{4\pi v^2}\,,\qquad \tilde \omega\equiv\omega/\mu\,,\qquad \tilde{\bf p}\equiv {\bf p}/\mu\,.\label{alpha}
\ee
We find,
\be
\tilde\omega_{\pm}\,=\,\sqrt{\tilde{\bf p}^2+(s+1)+\frac{1}{2\alpha^2}\pm\sqrt{4\tilde{\bf p}^2\,+\,\left(s+1-\frac{1}{2\alpha^2}\right)^2}}\,.
\ee
Both modes are gapped at $\tilde{\bf p}=0$. 
\bea
&&\tilde\omega^2_+\,\simeq\,2(s+1)+\tilde{\bf p}^2\,\frac{2\alpha^2(s+3)-1}{2\alpha^2(s+1)-1}\ldots\,, \\\nonumber \\\nonumber
&&\tilde\omega^2_-\,\simeq\,\frac{1}{\alpha^2}+\tilde{\bf p}^2\,\frac{2\alpha^2(s-1)-1}{2\alpha^2(s+1)-1}\ldots\,,\qquad\qquad |\tilde {\bf p}|\ll1\,.
\eea
When the Chern-Simons level is taken to be large the mode $\tilde\omega_{-}$ is the lighter of the two and becomes the phonon in the strict $k\to\infty$ limit.
For the classically marginal sextic potential with $s=3$, this limit yields $\tilde\omega_-^2\simeq \tilde {\bf p}^2/2$ which implies a speed of sound $c_s=1/\sqrt 2$, expected from scale invariance in 2+1 dimensions. As the level $k$ is lowered, the gaps of the two branches coincide when $\alpha=1/\sqrt{2(s+1)}$, and below this value of $\alpha$, the roots exchange roles.
\begin{figure}[h]
\begin{center}
    \includegraphics[width=0.55\textwidth]{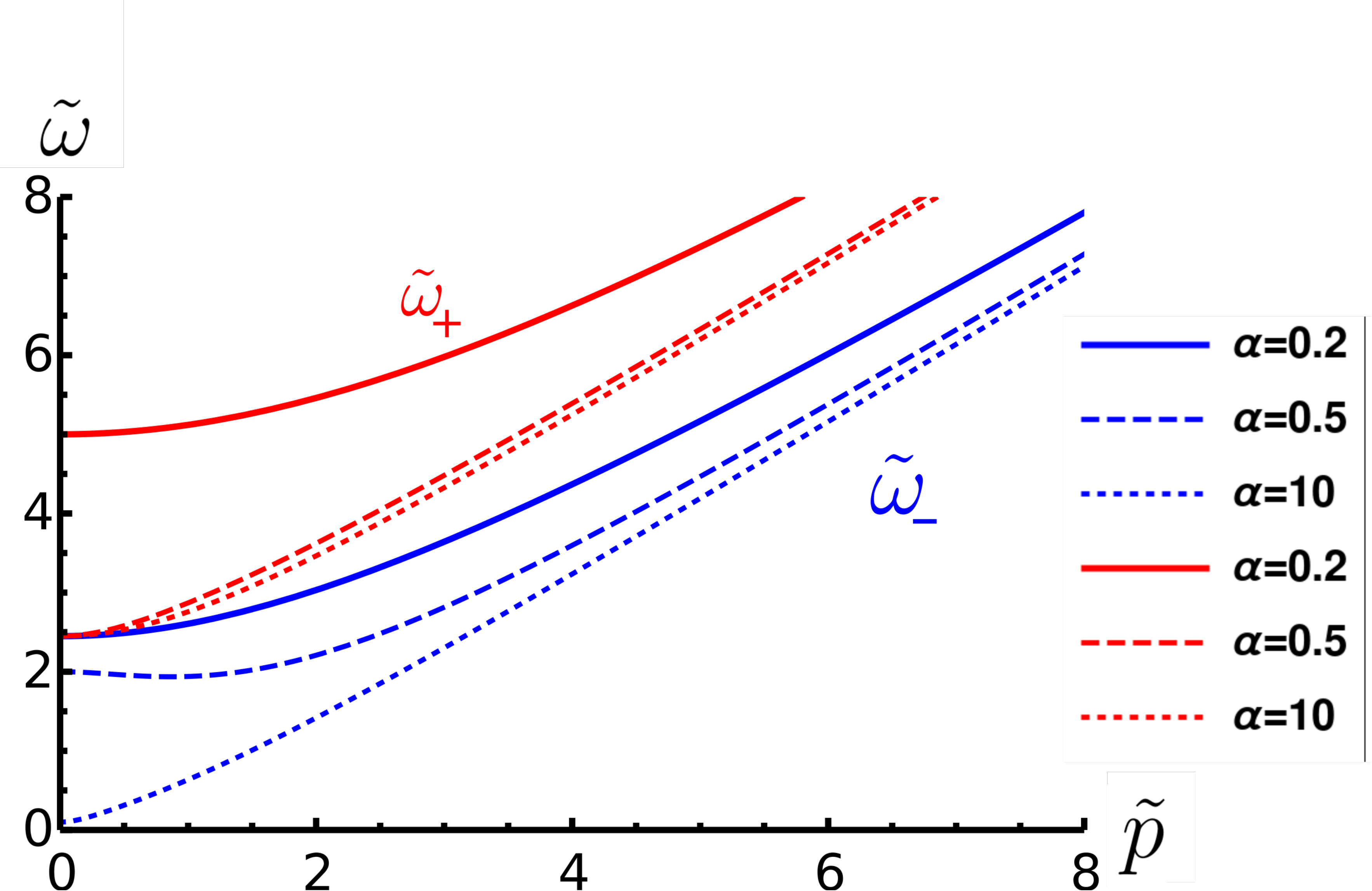}    
    \caption{{\small The dispersion relations of the two branches of perturbative fluctuations for the quartic potential ($s=2$), with different values of the dimensionless coupling $\alpha$.}} \label{fig:dispersion}
    \end{center}
\end{figure}
Depending on the value of $\alpha$, the sign of  $\tilde\omega''_{\pm}(0)$ can be negative which implies a minimum in the dispersion relation away from $\tilde{\bf p}=0$, also called a magneto-roton minimum. One  of the two branches  will exhibit a magneto-roton minimum  for values of $\alpha$ in the range:
\be
\quad\frac{1}{\sqrt{2(s+3)}} < \alpha <  \frac{1}{\sqrt{2(s-1)}}\,.
\ee

\section{Vortex equations}
We want to solve the equations of motion in polar coordinates. Therefore, allowing for a non-trivial spatial metric $h$, the static field equations are:
 \bea
        &&\tfrac{1}{\sqrt h}\partial_j \left(\sqrt h\,\partial^j\Phi\right) - \tfrac{i}{\sqrt{h}} \partial_{j} \left(\sqrt{h} A^{j} \Phi \right) + (\mu +A_0)^2 \Phi -i A_{j} \partial^{j} \Phi - A_{j} A^{j} \Phi\, =\, 
  g_{2s} |\Phi|^{2s-2}\Phi \nonumber\\\\\nonumber
        &&\frac{k}{2 \pi} \epsilon^{\sigma \nu \rho} \partial_{\nu} A_{\rho}\, = \,-\sqrt{h}\left[i \left(\partial^{\sigma} \Phi \Phi^{\dag} - \Phi \partial^{\sigma} \Phi^{\dag} \right) \,+\,2 A^{\sigma} \left|\Phi \right|^2- \delta_{0 \sigma} \left(2 \mu \left|\Phi \right|^2 + J_0 \right)\right]\,.
\eea
 As usual, a static vortex configuration carrying $n$ units of magnetic flux is described by the  rotationally symmetric ansatz in polar coordinates: 
\be
            A_0\, =\, A_0(r) \, , \qquad A_r \,=\, 0, \qquad A_{\theta}\, =\, A_\theta(r), 
\qquad \Phi \,= \,f(r)e^{i n \theta}\,.
        \ee
This ansatz leads to the following system of equations\footnote{We omit the equation of motion for $A_r$ which is automatically satisfied.}:
    \bea
&&        f''+\frac{f'}{r} -\frac{1}{r^2}\left(A_\theta - n\right)^2 f + (A_0+\mu)^2 f\, - g_{2s} f^{2s-1} = 0 \label{eq:radial_scalar}  \\\nonumber\\
 &&       -\frac{f^2}{r}\left(A_\theta - n\right)+ \frac{k}{4 \pi} A_0' =0 \label{eq:radial_Atheta}  \\\nonumber\\
 &&       -f^2 A_0 -\mu f^2 + \frac{k}{4 \pi} \left(\frac{A'_\theta}{r}\right)\,=\, \frac{J_0}{2}\,,\qquad \qquad J_0\,=\,-\,2\mu\,v^2\,. \label{eq:radial_A0}
    \eea
We are looking for a configuration which asymptotes at infinity to the ground state with the fixed non-vanishing background value for $J_0$, obeying the boundary conditions:
        \bea
            A_\theta(r) \xrightarrow{r \to \infty} n\,, \qquad A_0(r) \xrightarrow{r \to \infty} 0, \qquad f(r) \xrightarrow{r \to \infty} v\,, \qquad f(0) = 0\,.
        \eea
With the rotationally symmetric ansatz above, the magnetic field $B$ only depends on $r$, and the configuration carries $n$ units of flux:
\be
B(r)\,=\,\frac{A_\theta'(r)}{r}\,,\qquad\qquad
\Phi_B 
\,=\,\lim_{r\to\infty} \int_0^{2 \pi} A_{\theta}(r)\, d\theta \,=\, 2\pi n\,,
\ee
assuming  $A_\theta$ vanishes at the origin.
\subsection{Qualitative features}
The equation of motion \eqref{eq:radial_A0} for $A_0$, which implements the Gauss constraint, fixes the value of the magnetic field at the core of the vortex where the scalar field vanishes. Thus,
\be
B(0)\,=\,\frac{4\pi}{k}\,J_0
\,=\,-\frac{\mu^2}{\alpha}\,,
\ee
where $\alpha$ is the effective coupling defined in \eqref{alpha}. For a given winding number $n$, the value of $\alpha$ completely characterizes the vortex solution when expressed in terms of rescaled dimensionless variables.
The effect of the Gauss constraint is to induce a magnetic field $B$ inside the vortex core to precisely cancel the background source $J_0$ which could also  be viewed as a uniform background magnetic field.  Outside the vortex $B$ vanishes,
\be
\lim_{r\to\infty} B(r)\,=\,0\,.
\ee

The sign of $B(0)$ plays an important role in determining the properties of the vortex solutions. Without loss of generality, we assume that the chemical potential $\mu$ and the Chern-Simons level $k$ are both positive:
\be
\mu >0\, \qquad \qquad k >0\,.
\ee
With this choice $B(0)$ is negative definite,   independently of the sign of the magnetic flux $2\pi n$. However, this means that solutions with positive and negative flux will be qualitatively different. This breaking of charge conjugation is precisely what we expect in the presence of the $U(1)$ chemical potential.

\paragraph{Negative flux $n<0$}: Assuming that the magnitude of the magnetic field $|B(r)|$ increases monotonically towards the vortex core, and given that the value of the core magnetic field is independent of $|n|$ (the number of units of flux) the vortex core size should  increase with $|n|$ for negative $n$. Taking $B$ to be uniform within the core for large enough $|n|$, we can estimate the radius $R_n$ of a vortex solution with $|n|\gg 1$:
\be
|B(0)|\pi R_n^2 \,\approx \, 2\pi |n| \,\quad\implies\quad R_n\,\approx\,\frac{\sqrt{2\alpha|n|}}{\mu}\,,\quad |n|\gg1\,.
\label{radius}
\ee
The assumption of uniformity of the vortex core region is self-consistently justified by first noting that for small $r$,
\be
f(r)\,=\,c_0\,r^{|n|} +\ldots\,,\qquad \qquad c_0 >0\,.
\ee 
This is obtained by neglecting $A_\theta$ in comparison to $n$, and ignoring higher order terms for small $r$ in eq.\eqref{eq:radial_scalar}. 
Therefore, for large flux, the scalar profile is extremely flat near $r=0$. With a uniform $B$ field in the vortex interior, the vector potential is determined as 
\be
 A_\theta\approx-\frac{1}{2} |B(0)| r^2\,,
\ee
and this approximation breaks down precisely when $r\approx R_n$
(see eq.\eqref{radius}) . 
We see numerically that the scalar field profile,  inside the vortex, closely follows the solution to the first order equation:
\bea
&&f'(r)\,\approx\,\tfrac{1}{r}\left(|n|-\tfrac{1}{2}|B(0)|r^2\right)\, f(r)\\\nonumber\\\nonumber
\implies \quad&&f(r)\approx c_0 r^{|n|}\,e^{-|B(0)|r^2/4}\,,\quad r< R_n\,.
\eea
This feature of the solution  is depicted in figure \ref{fig:firstorder}. 

The equation of motion \eqref{eq:radial_Atheta} for $A_\theta$ determines the  radial electric field $E(r)\,=\,A_0'(r)$. 
The electrostatic potential $A_0$ remains constant inside the core region since $f(r)$ is vanishingly small and therefore the electric field is also vanishingly small. 
Outside the core region the electric and magnetic fields decay exponentially to zero. Linearizing about the asymptotic solution at  large $r$ we find,
\bea
&&\delta f\,\equiv\, f(r)-v\,,\qquad \delta A_\theta\,=\, A_\theta-n\,,\\\nonumber\\\nonumber
&&\delta A_\theta\,=\,\frac{\alpha}{\mu}\, r A_0'\,,\\\nonumber\\\nonumber
&&\delta f'' \,+\,\frac{\delta f'}{r}\,-\,(2s-2)\mu^2\delta f\,=\, -2 A_0 \mu v\,,\\\nonumber\\\nonumber
&&A_0''\,+\,\frac{A_0'}{r}\,-\,\frac{\mu^2}{\alpha^2}A_0\,=\,\frac{2\mu^3}{v\alpha^2}\delta f\,.
\eea
The solutions to the homogeneous equations for the fluctuations $\delta f$ and $A_0$ are the Bessel functions $K_0\left(\sqrt{2s-2}\mu r\right)$ and $K_0\left(\mu r/\alpha\right)$ respectively, and these control the exponential decay of fluctuations at large $r$,
\be
K_0\left(\sqrt{2s-2}\,\mu r\right)\big|_{\mu r\to\infty} \,\sim\,\frac{e^{-\sqrt{2s-2}\,\mu r}}{\sqrt {\mu r}}\,,\qquad K_0\left(\mu r/\alpha\right)\big|_{\mu r\to\infty} \,\sim\,\frac{e^{-\mu r/\alpha}}{\sqrt {\mu r}} \,.         
\ee
The two exponents are related to  masses of the gapped perturbative excitations around the Higgsed ground state.
Since $A_\theta(r)\simeq n$ outside the core region, according to  eq. \eqref{eq:radial_Atheta} the electric field is only significant in a strip  along the edge of the vortex core.  Numerically, we find that the width of this strip does not scale with $|n|$, so that in the limit of large $|n|$, the contribution from the transition region to the vortex energy is subleading in $|n|$. 

We will see below that these qualitative aspects of the vortex solutions lead to linear dependence of the vortex energy on $|n|$ and BPS-like behaviour at a critical value of the effective coupling $\alpha$.

\paragraph{Positive flux $n>0$:} Positive flux solutions are qualitatively distinct from the negative flux ones.  This is because $B(0)<0$ independent of $n$, so 
$B(r)$ must switch sign to yield a net positive flux.  For large $n>0$, most of this positive flux remains concentrated in a ring-like region at the edge of the vortex. In this case the total flux can be written as a sum of two contributions, one that scales with the area of the vortex and is negative, therefore must be subleading, and a positive dominant contribution which scales with the radius of the configuration\,,
\be
2\pi n\,\sim\, -\pi R_n^2\frac{\mu^2}{\alpha}\,+\, 2\pi R_n \Delta_{\rm ring} \, B_{\rm ring}\,.\label{ringansatz}
\ee
Here $\Delta_{\rm ring}$ is the width of the  edge region which we take to be independent of $n$, whilst 
$B_{\rm ring}$
denotes the peak value of the magnetic field in the ring and $R_n$ 
is the radius of the vortex for large enough $n$. The negative area-dependent contribution  is a key difference from a similar situation discussed in \cite{Bolognesi:2007ez}.  We cannot exclude the possibility that both the area and perimeter 
contributions have faster than linear growth and a delicate cancellation yields the correct flux. It is possible to suppress the area contribution by making $\alpha$ arbitrarily large. The large $\alpha$ limit makes the magnetic field inside the vortex vanishingly small and the system then closely resembles an abelian  Chern-Simons vortex in vacuum.  
\section{Vortex energy and BPS-like scaling}
The ``energy" functional appropriate for the grand canonical ensemble is the grand potential. The Chern-Simons term does not contribute to it for static configurations. The grand potential is obtained by using the Lagrangian density \eqref{lagfull} with $\mu\neq 0$ and passing to the Hamiltonian picture. Rewriting the Hamiltonian in terms of the fields and their derivatives, the desired energy functional is\footnote{For static configurations, the grand potential can be quickly derived by retaining only the spatial gradient and potential terms including the Chern-Simons density,
 \bea
\mathcal{E} \,=\,\int d^2 x 
 \left( \left| D_i \Phi\right|^2 \,- \,\left(A_0+\mu\right)^2 \left| \Phi\right|^2  
 + A_0\left(\frac{k}{2 \pi}  B -J_0\right)+ \frac{g_{2s}}{s}  \left| \Phi\right|^{2s}\right)\,,
 \eea
 and then applying the constraint \eqref{constraint}.
},
\bea
{\cal E}\,=\,\int d^2 x 
 \left( \left| D_i \Phi\right|^2 \,+ \,A_0^2 \left| \Phi\right|^2  
 +\frac{g_{2s}}{s}  \left| \Phi\right|^{2s}-\mu^2\left| \Phi\right|^2\right)\,,
 \eea 
 accompanied by the constraint which incorporates the effect of the Chern-Simons term,
 \be
\frac{k}{4\pi} B\,=\,A_0|\Phi|^2+\mu\left(|\Phi|^2-v^2\right)\,.\label{constraint}
 \ee
 
An interesting feature of the finite-$\mu$ vortices (with negative flux)  is that they appear to be  marginally bound (or BPS-like) for a specific value of the effective dimensionless parameter $\alpha$. Recall that this parameter depends on the Chern-Simons level, the chemical potential and the interaction strength:
\be
\alpha \,=\,\frac{k\mu}{4\pi v^2}\,=\,\frac{k}{4\pi}\left(\frac{g_{2s}}{\mu^{3-s}}\right)^{\frac1{s-1}}\,.\label{alphadef}
\ee
Let us perform a rescaling of variables and fields so that the equations of motion can be written explicitly in terms of dimensionless quantities:
\be
\tilde r\,\equiv\,\mu r\,,\qquad \tilde f\,\equiv\, \frac{f}{v}\,,\qquad a\,\equiv\,\frac{1}{\tilde r} \left(A_\theta-n\right)\,,\qquad \tilde A_0\,\equiv\,\frac{A_0}{\mu}\,.
\ee
The rescaled scalar profile vanishes at  $\tilde r=0$ and approaches unity for large $\tilde r$: $\tilde f(\tilde r\to\infty)=1$.  The asymptotic behaviours of $a(\tilde r)$ and $\tilde A_0$ are,
\be
\lim_{\tilde r\to 0}\tilde r a(\tilde r)\,=\,-n\,,\qquad \lim_{\tilde r\to \infty}\tilde r a(\tilde r)\,=\,0\,,\qquad \lim_{\tilde r\to \infty}\tilde A_0(\tilde r)=0\,.
\ee
The resulting dimensionless equations of motion (primes denote derivatives with respect to $\tilde r$) are,
 \bea
 &&\tilde f'' + \frac{\tilde f'}{\tilde r} - a^2\, \tilde{f} + \left(\tilde A_0+1\right)^2\tilde{f}  -\tilde f^{2s-1} =0 \label{eq:dimlesseom1}\\\nonumber\\
&& \alpha \tilde A_0' \,=\,  a\,\tilde f^2 \label{eq:dimlesseom2}\\\nonumber\\
  &&      \alpha \tilde B= \left(\tilde f^2-1\right)+\tilde f^2 \tilde A_0\,.\label{eq:dimlesseom3}
 \eea
Therefore, for a fixed flux $n$, the solutions are only parametrised by the  dimensionless 
effective coupling $\alpha$. Note we have introduced the
dimensionless magnetic field,
\be
\tilde B\,=\,\frac{1}{\tilde r}\,\left(\tilde r a\right)^\prime\,=\,\frac{A_\theta^\prime}{\tilde r}\,.
\ee
We first rewrite the energy functional using the rescaled fields and variables,
\be
{\cal E}\,=\,2\pi v^2\int_0^\infty \tilde r\, d\tilde r\left[\left(\tilde f'- a \tilde f\right)^2 + a\frac{d }{d\tilde r}\left(\tilde f^2-1\right)+\tilde A_0^2\tilde f^2 +\frac{\tilde f^{2s}}{s}-\tilde f^2 +\frac{s-1}{s}
\right]\,,\label{bpsfirst}
\ee
where we have included a constant zero-point shift so that the energy density is vanishing for the ground state at infinity. The second term in eq.\eqref{bpsfirst} when integrated by parts yields a nonvanishing surface contribution,
\be
\int_0^\infty d\tilde r \,\left(\tilde r a\right)\frac{d }{d\tilde r}\left(\tilde f^2-1\right)\,=\,|n|\,-\,\int_0^\infty \tilde r\, d\tilde r \tilde B(\tilde f^2-1)\,.
\ee
Employing the Gauss constraint to eliminate $\tilde B$ in favour of $\tilde A_0$, we obtain an expression for the energy functional which is suitable for subsequent approximations and matching with numerical results,
\bea
{\cal E} = 2\pi v^2 |n| +2\pi v^2\int_0^\infty \tilde r d\tilde r&&\left[\left(\tilde f'-a\tilde f\right)^2+\left(\frac{\tilde f^{2s}}{s}-\tilde f^2 +\frac{s-1}{s}\right)-\frac{1}{\alpha}\left(1-\tilde f^2\right)^2 +\right.\nonumber\\\label{energys}\\\nonumber
&&\left.\tilde A_0^2\,\tilde f^2+\frac{1}{\alpha}\tilde A_0 \tilde f^2(1-\tilde f^2)\right]\,. 
\eea
 \subsection{The quartic potential $s=2$}
Anticipating numerical results in the next section, we can make certain  observations on the energetics of vortex solutions for large negative flux. 

Our arguments rely on the fact that for $n$ sufficiently large and negative, all fields are uniform inside and outside the vortex whose radius scales as $\sqrt{|n|}$ , with a thin transition region whose width does not scale with $n$. In the case of the quartic potential the energy functional is, 
\bea
{\cal E}(n, \alpha)\big|_{s=2} = 2\pi v^2 |n| +2\pi v^2\int_0^\infty \tilde r d\tilde r&&\left[\left(\tilde f'-a\tilde f\right)^2+\left(\tfrac{1}
{2}-\tfrac{1}{\alpha}\right)\left(1-\tilde f^2\right)^2 +\right.\nonumber\\\label{energys2}\\\nonumber
&&\left.+\tilde A_0^2\,\tilde f^2+\tfrac{1}{\alpha}\,\tilde A_0 \tilde f^2(1-\tilde f^2)\right]\,. 
\eea
We know that $\tilde f$ vanishes inside the vortex, whilst $\tilde A_0$  and $a(r)$ vanish outside it.  Specifically when $\alpha=2$, the scalar potential is precisely cancelled, and the integrand in the expression above has support only within the transition region at the edge of the vortex. If we assume that this  contribution does not scale with $n$, we conclude that
\be
{\cal E}(n, \,\alpha=2)\big|_{s=2, |n|\gg1}\,=\, 2\pi v^2 |n|\,.
\ee
Our numerical solutions confirm (figure \ref{fig:massn1}) this conclusion which works remarkably well even for low values of $n$, including  $|n|=1,2 \ldots$. Another surprising feature of the solutions for $\alpha=2$ is that they appear to solve the first order equation $\tilde f'= a \tilde f$ to very high numerical accuracy (figure \ref{fig:firstorder})\footnote{In this context, it is worth noting that if the terms proportional to $\tilde A_0$ are omitted (or assumed to be negligible) in eqs. \eqref{eq:dimlesseom1} and \eqref{eq:dimlesseom3}, then the resulting equations coincide, for a particular value of $\alpha$, with equations of motion for a BPS vortex in $U(2)\times U(2)$ ABJM theory which solves an equivalent first order system \cite{Auzzi:2009es}.}.

It is easy to extend the arguments to  other values of $\alpha$. When $\alpha\neq 2$, the second term in the integrand of \eqref{energys2} provides an approximately constant energy density inside the vortex where $\tilde f=0$, while all other contributions remain vanishingly small. We therefore find,
\be
{\cal E}(n, \,\alpha)\big|_{s=2, |n|\gg1}\,=\, 2\pi v^2 \left(|n|\,+\,\tfrac12 \mu^2R_n^2\left(\tfrac12-\tfrac{1}{\alpha}\right) \right)\,=\,\alpha\pi v^2|n|\,.
\ee
This behaviour is also confirmed numerically in figure \ref{fig:massn1}, albeit only for large $|n|$ as expected. Interestingly, although $v$ and $\alpha$ each  depend on the quartic coupling constant $g_4$, the large-$|n|$ vortex mass formula depends only the Chern-Simons level $k$ and the chemical potential,
\be
{\cal E}(n,\,\alpha)\big|_{s=2, |n|\gg1}\,=\,\alpha\pi v^2|n|\,=\, \frac{k\mu}{4}|n|\,.
\ee

\subsection{General power law potential $(s\geq 2)$}
 The energies of solutions for general power law potentials now work along similar lines.  The integrand in the energy density \eqref{energys} is negligible both inside and outside the vortex when $\alpha=\alpha_c$,
 \be
 \alpha_c= \frac{s}{s-1}.
 \ee 
 At this critical  $\alpha$ we expect solutions with large flux to have energies ${\cal E}(n)\simeq 2\pi v^2 |n|$. When $\alpha$ takes generic values away from  $\alpha_c$, adapting the $s=2$ argument to general power law potentials $\sim g_{2s}|\Phi|^{2s}$  we obtain,
 \be
 {\cal E}(n, \,\alpha; s)\big|_{|n|\gg 1}\,=\,2\alpha\pi v^2|n|\frac{s-1}{s}\,=\,\frac{s-1}{2s}k\mu|n|\,.\label{generalenergy}
 \ee
This result is confirmed by our numerical solutions for the sextic potential below. As before the mass formula is independent of the interaction strength of the potential. We point out however that the radius of the vortex solution $R_n=\frac{1}{\mu}\sqrt{2\alpha |n|}$ depends nontrivially on all parameters. From the definition of $\alpha$ \eqref{alphadef}, increasing the interaction strength (for fixed $\mu$ and $k$) has the effect of increasing the vortex size.
\subsection{Positive flux vortices}
With our choice of conventions  $\mu >0$ and $k>0$, positive flux solutions are energetically disfavoured. To see this, let us  reconsider the energy functional \eqref{bpsfirst}.  As in the case of the flux in eq.\eqref{ringansatz}, there are different types of contributions to the energy, those that scale with the area of the vortex, those that scale with the perimeter, and finally (gauge-covariant) gradient  terms which become important at the edge of the vortex.
If we pick $\alpha$ to be sufficiently large so that we can ignore the flux contribution from the vortex interior, then 
$B_{\rm ring}\sim n/R_n$. Then the leading contributions to the energy take the schematic form,
\be
{\cal E}(n)\sim (2\pi v^2)\left(\frac{n^2}{R_n}\Delta + c_0\pi R_n^2 \right)\,,\qquad \alpha\gg1
\ee
Here the first term is a perimeter contribution from a covariant gradient while the second term is a potential energy contribution from the interior. Extremizing with respect to $R_n$, we obtain $R_n\sim n^{2/3}$ and ${\cal E}(n)\sim n^{4/3}$.
However this argument will fail for  finite $\alpha$ since $B(0)=-\mu^2/\alpha$ and the area scaling as $n^{4/3}$ would violate  the flux condition \eqref{ringansatz}.  We have not found a satisfactory scaling argument for finite $\alpha$ and large $n$, but our numerical results indicate a faster than quadratic growth of the energy as a function of $n$ in this situation.

%

\section{Numerical results}
The vortex equations of motion are not analytically solvable. We numerically solve the dimensionless equations of motion \eqref{eq:dimlesseom1}, \eqref{eq:dimlesseom2} and \eqref{eq:dimlesseom3} along with the accompanying boundary conditions. Below we outline the results for the quartic potential $(s=2)$ first, and subsequently  summarize the results for the sextic $(s=3)$ case.
\subsection{Quartic potential $(s=2)$}
\begin{figure}[h]
\begin{center}
    \includegraphics[width=3in]{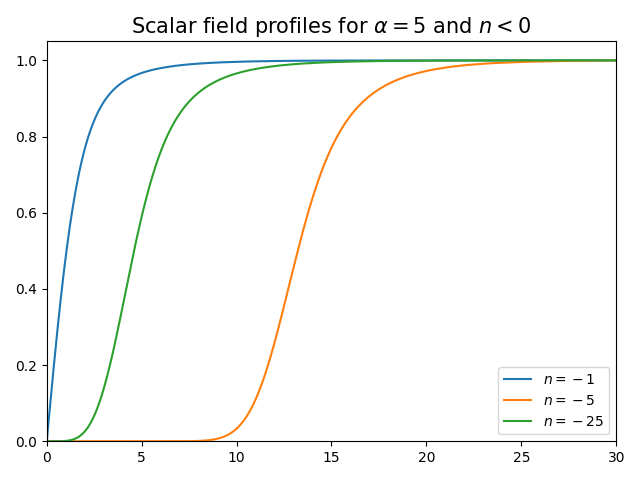}\includegraphics[width=3in]{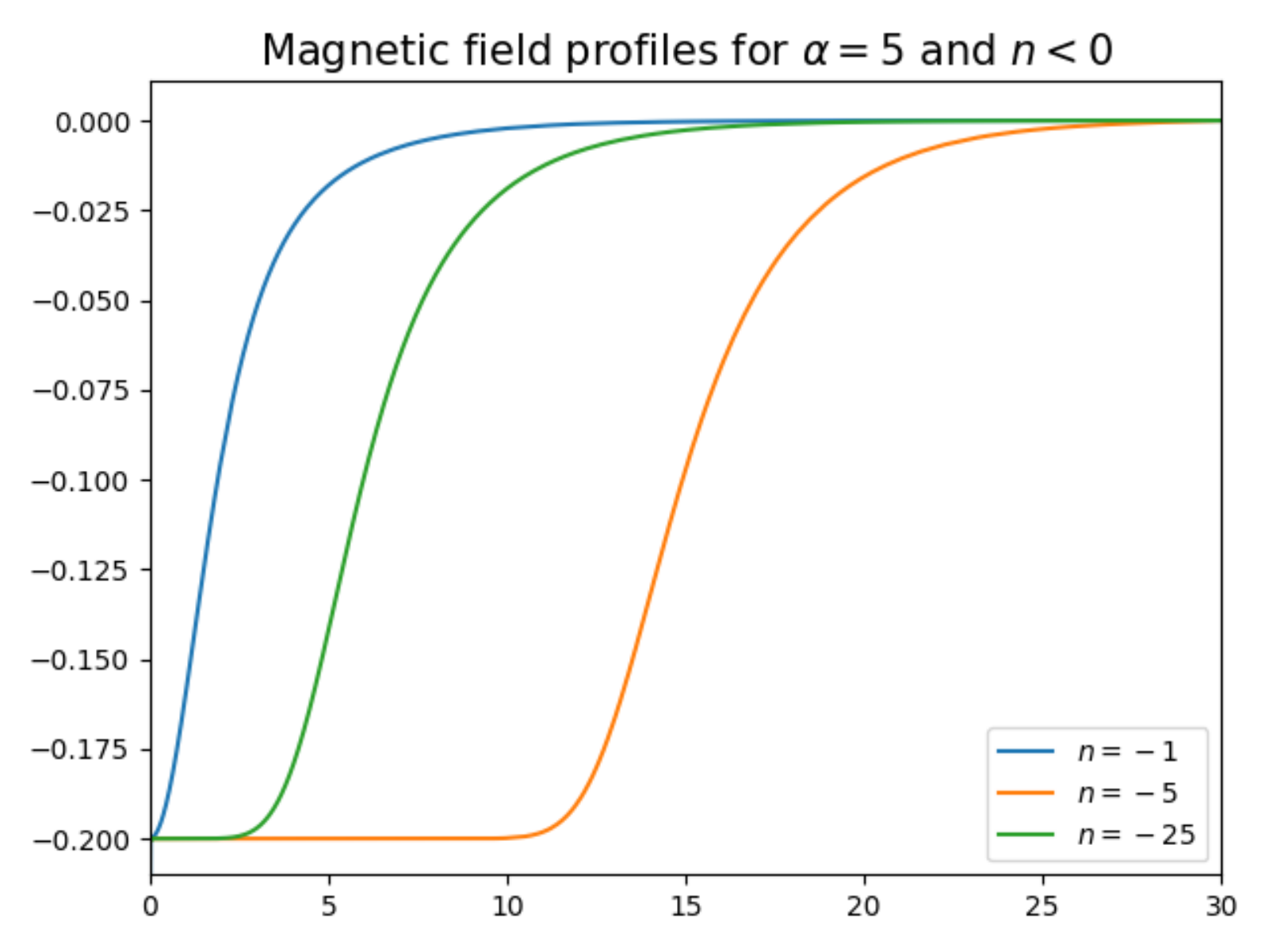}
   \caption{ {\small{\it Left}: The scalar field profile $\tilde f(\tilde r)$ for $\alpha=5$ and negative flux. {\it Right:} The dimensionless magnetic field $\tilde B(\tilde r)$ for the same values of $\alpha$ and magnetic flux.}} \label{fig:scalarneg}
   \end{center}
\end{figure}

\paragraph{Negative flux solutions:} A well known feature of Chern-Simons-Higgs vortices in vacuum is the ring-like profile of electric and magnetic fields \cite{Horvathy:2008hd}. The finite density vortices we have studied are qualitatively distinct and
retain this feature only partially. Figure  \ref{fig:scalarneg}  shows the (dimensionless) scalar field $\tilde f(\tilde r)$ and magnetic field $\tilde B(\tilde r )$  at $\alpha=5$ and different negative values of the magnetic flux.
Unlike Chern-Simons vortices in vacuum \cite{Horvathy:2008hd, Bolognesi:2007ez} the magnetic field is no longer expelled from the core of the vortex. Instead, the magnetic field and the scalar are both effectively constant separately inside and outside the vortex and we observe a kink-like transition in between.  The value of the dimensionless magnetic field inside the vortex is $\tilde B(0)= -1/\alpha=-0.2$ for $\alpha=5$. 

\begin{figure}[h]
\begin{center}
    \includegraphics[width=2.8in]{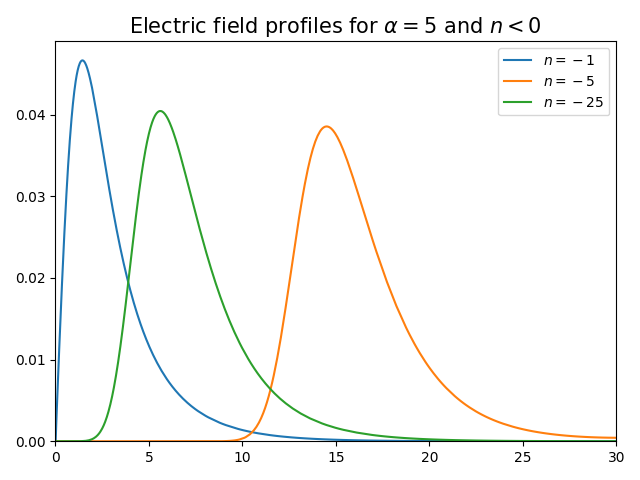} \includegraphics[width=2.8in]{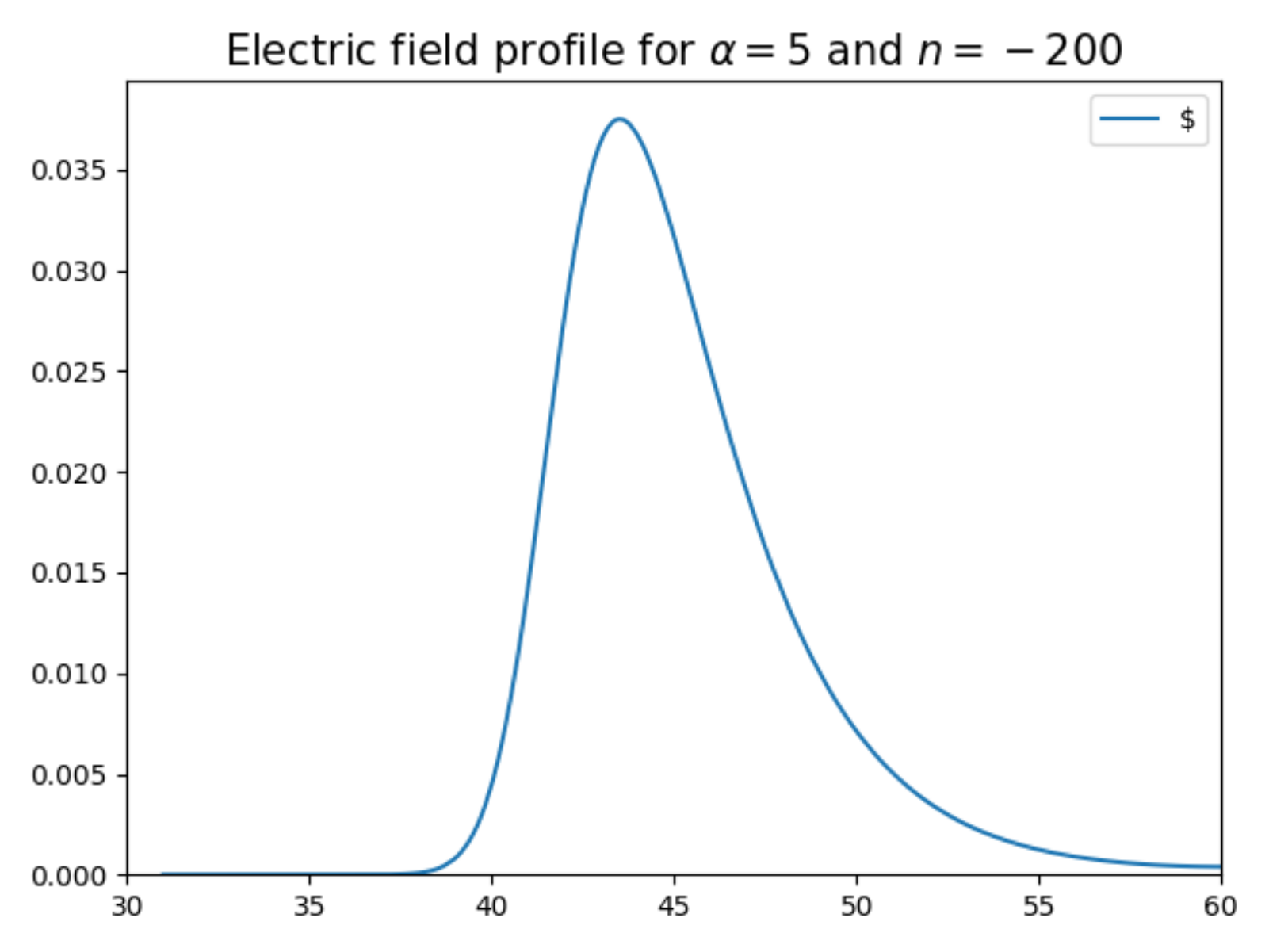} 
   \caption{ {\small The electric field has support only at the edge of the vortex implying a ring-like profile. 
   The width of the transition region remains fixed as $|n|$ is cranked up.} }\label{fig:Eneg}
   \end{center}
\end{figure}

 The electric field on the other hand has support only at the edge or the transition-region where the gradient of $A_0$ is significant. This is illustrated in figure \ref{fig:Eneg}.
Even for relatively low values of $|n|$, the location of the peak in the magnitude of the electric field begins to track the large $|n|$ estimate of the vortex radius \eqref{radius}:
\bea
\mu R_n = \sqrt{2|n|\alpha} = \quad
\left\{\begin{matrix}
7.07 & \quad n=-5\,,\alpha=5\\
15.81&\quad n=-25\,,\alpha=5\\
44.7 & \quad n=-200\,,\alpha=5\,.
\end{matrix}\right.
\eea
The width of the ring-like transition region remains fixed as $|n|$ is increased. In particular, both the width of the ring and the peak magnitude of the electric field appear to be solely determined by $\alpha$ for large flux.
\begin{figure}[h]
\begin{center}
    \includegraphics[width=2.9in]{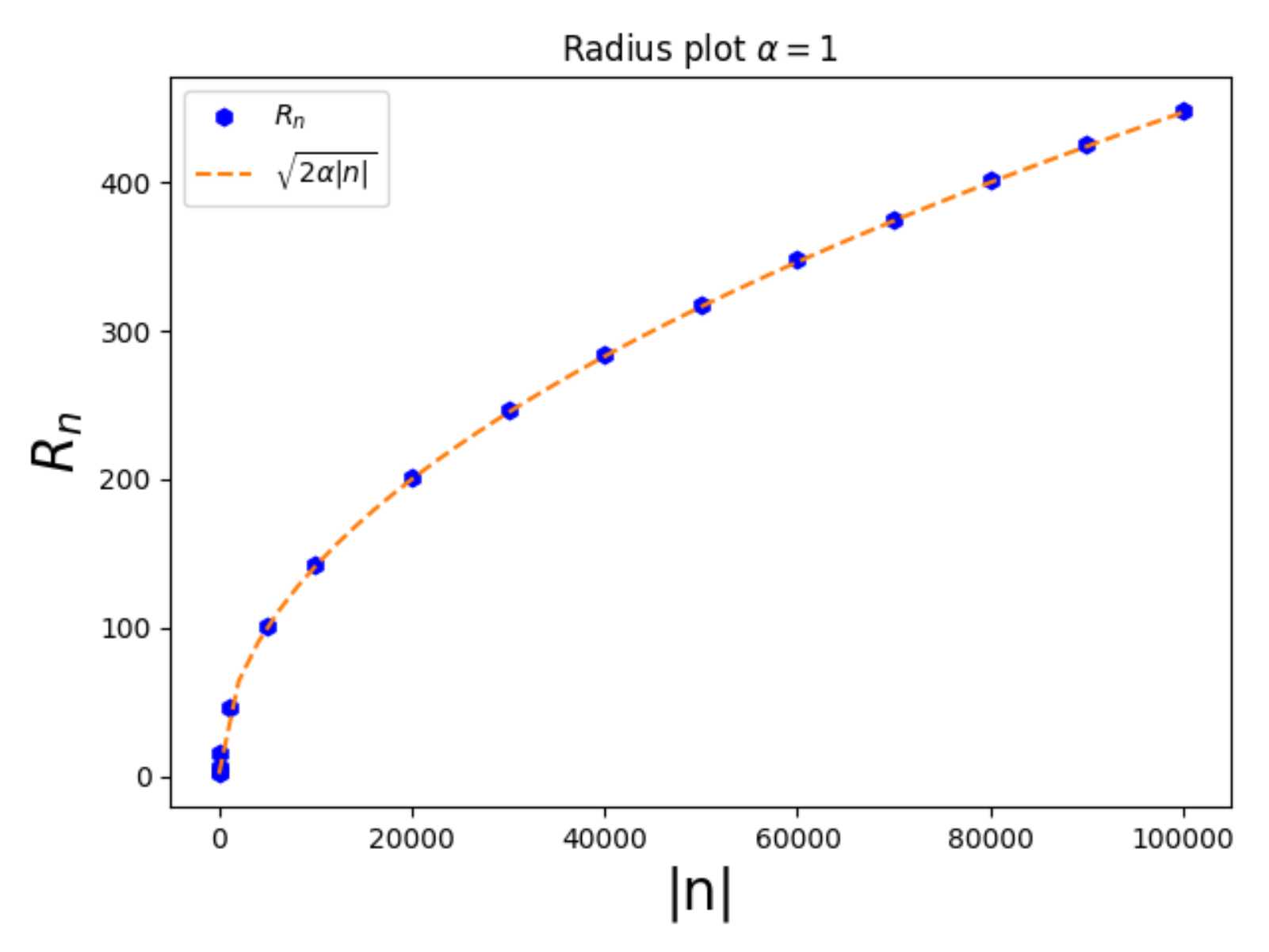}  
   \caption{ {\small The radii of large-$n$ vortex solutions follow very closely the curve $\mu R_n=\sqrt{2\mu |n|}$.
We define the radius of the vortex as the position at which the dimensionless energy density falls below a threshold $\sim 10^{-4}$. } }\label{fig:radius}
   \end{center}
\end{figure}
A precise agreement between the above scaling formula for the radius is obtained when the magnitude of the flux is significantly increased, as shown in the second plot in figure \ref{fig:radius}.

The most interesting aspect of the negative flux vortex solutions is the scaling of the energy with $|n|$. Using the energy functional \eqref{bpsfirst}, we compute the two dimensionless ratios,
\be
\frac{{\cal E}(n, \alpha)}{{|n|\cal E}(1, \alpha)}\,\qquad{\rm and} \qquad \frac{{\cal E}(n, \alpha)}{2\pi v^2|n|} \quad\xrightarrow{|n|\gg 1} \quad\frac{\alpha}{2}\,.\label{ratios}
\ee
The first ratio measures the energy of the $n$-vortex relative to that of $|n|$ vortices each with unit (negative) flux. If this is less than unity, then the $n$-vortex has lower energy than $|n|$ separated $-1$-vortices, and therefore the interactions between them must be attractive (type I). Conversely, if ${\cal E}(n, \alpha) > |n|{\cal E}(1, \alpha)$, the vortex interaction is repulsive (type II). 
\begin{figure}[h]
\begin{center}
 \includegraphics[width=2.90in]{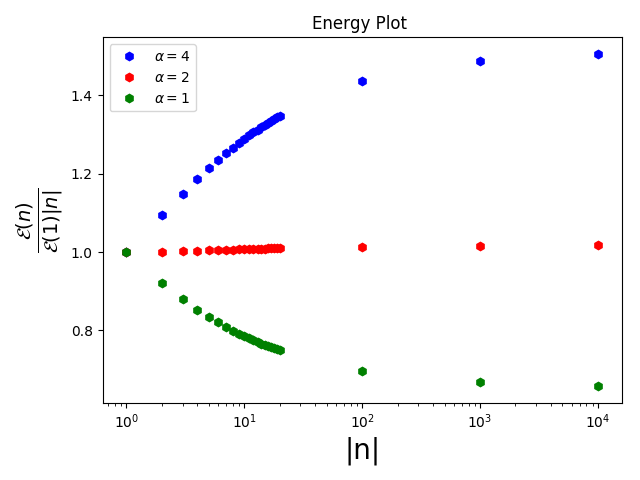}\hspace{0.1in}
    \includegraphics[width=2.90in]{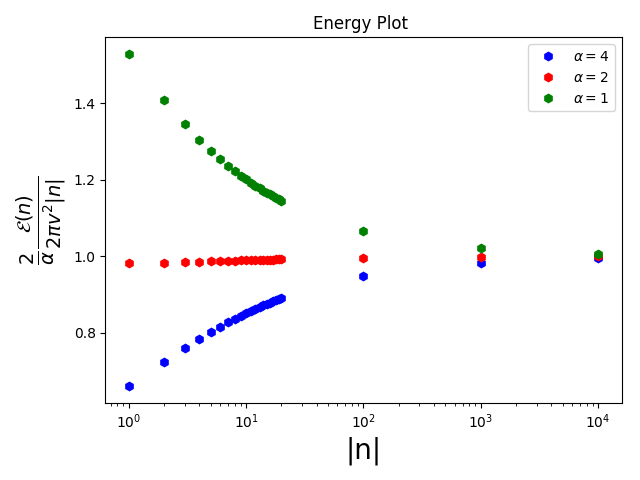}
   \caption{{\small{\it Left:} This shows that $\alpha < 2$ vortices are type I (attractive), separated from type II (repulsive) solutions for $\alpha>2$ by the $\alpha=2$ line where the solutions have vanishing interaction energy. {\it Right:} The $\alpha$-dependence of the $n$-vortex mass formula agrees with analytical arguments at large $|n|$. }} \label{fig:massn1}
   \end{center}
\end{figure}
The second ratio in \eqref{ratios} is the general formula for the $\alpha$-dependence of the $n$-vortex energy which was deduced from arguments for large $|n|$. Figure \ref{fig:massn1} shows to significant numerical precision that negative flux solutions with $\alpha=2$ are effectively ``BPS" for any value of  $|n|$, separating $\alpha> 2$ solutions which are type II (repulsive) from the solutions with $\alpha < 1$ which are type I or attractive. Furthermore the $\alpha$-dependence of the energies of type I and type II vortices for large flux matches the predicted behaviour in eq.\eqref{ratios}. 

\begin{figure}[h]
\begin{center}
    \includegraphics[width=2.80in]{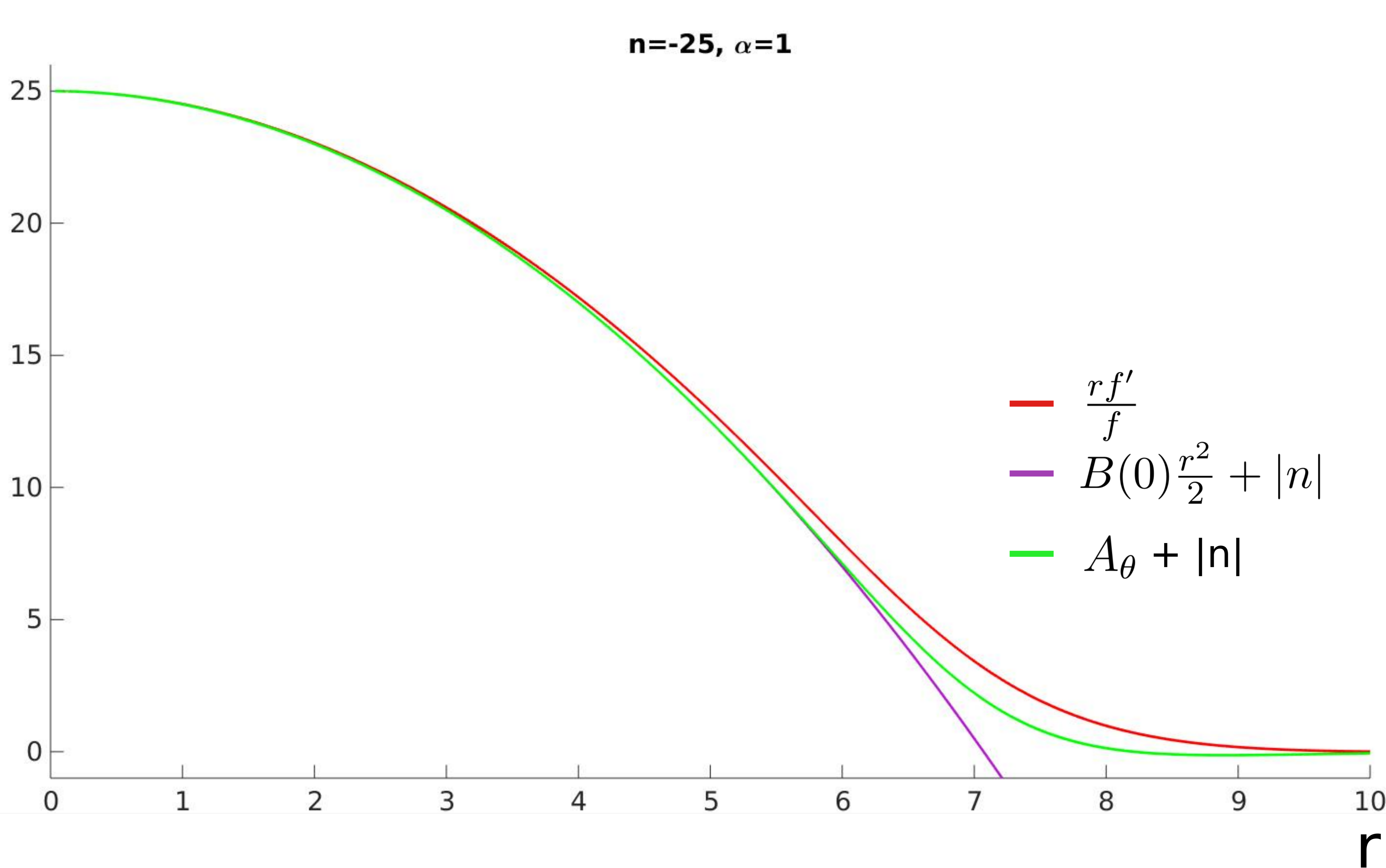} \hspace{0.1in}
      \includegraphics[width=2.8in]{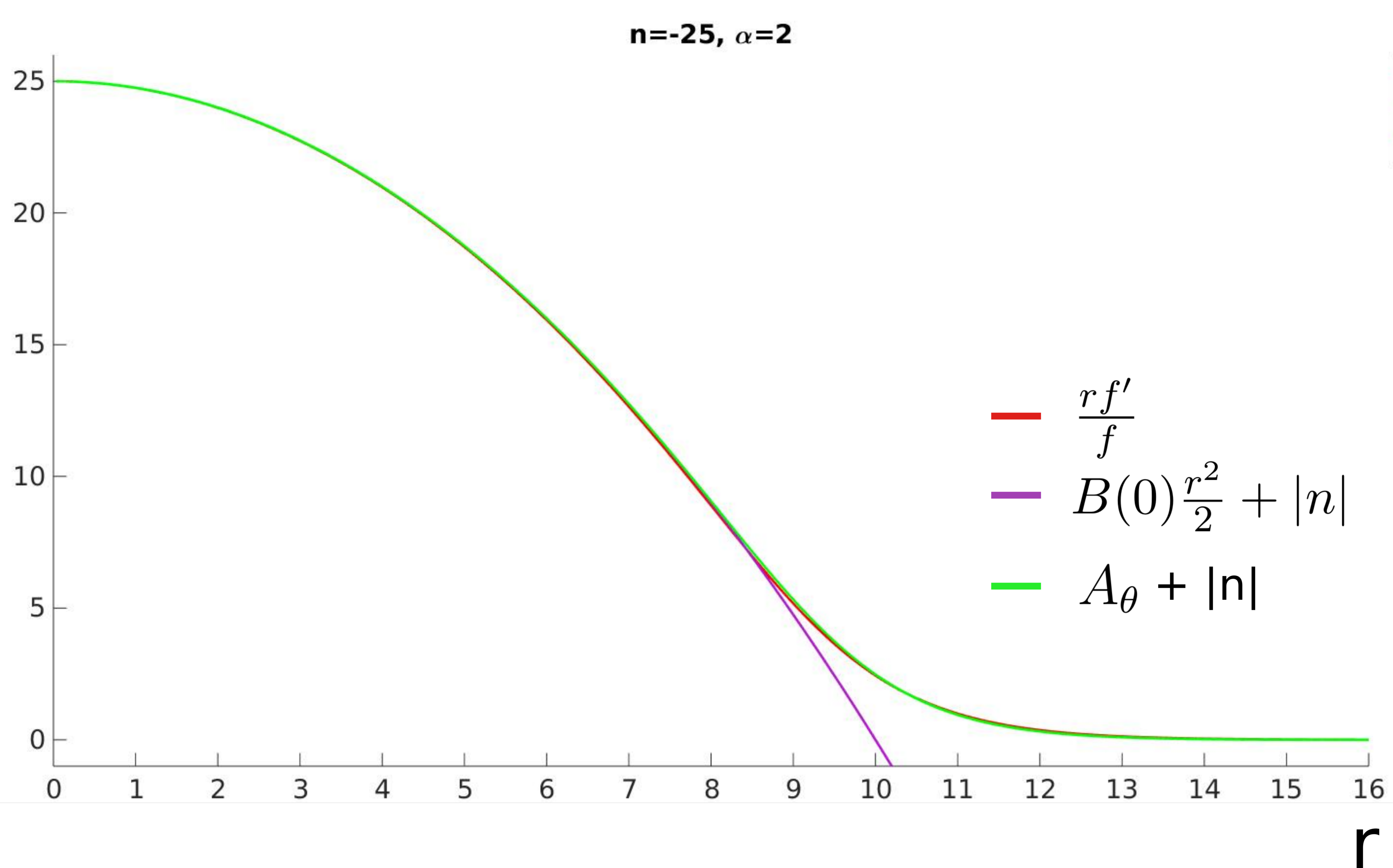} 
    \caption{{\small The figure shows the three quantities: $rf'/f$ (red), $\left(A_\theta+|n|\right)$ (green), and $\left(-|B(0)|r^2/2+|n|\right)$ (purple) evaluated on the exact numerical solutions for $\alpha=1$ and $\alpha=2$.}} \label{fig:firstorder}
    \end{center}
\end{figure}

\begin{figure}[h]
\begin{center}
    \includegraphics[width=2.7in]{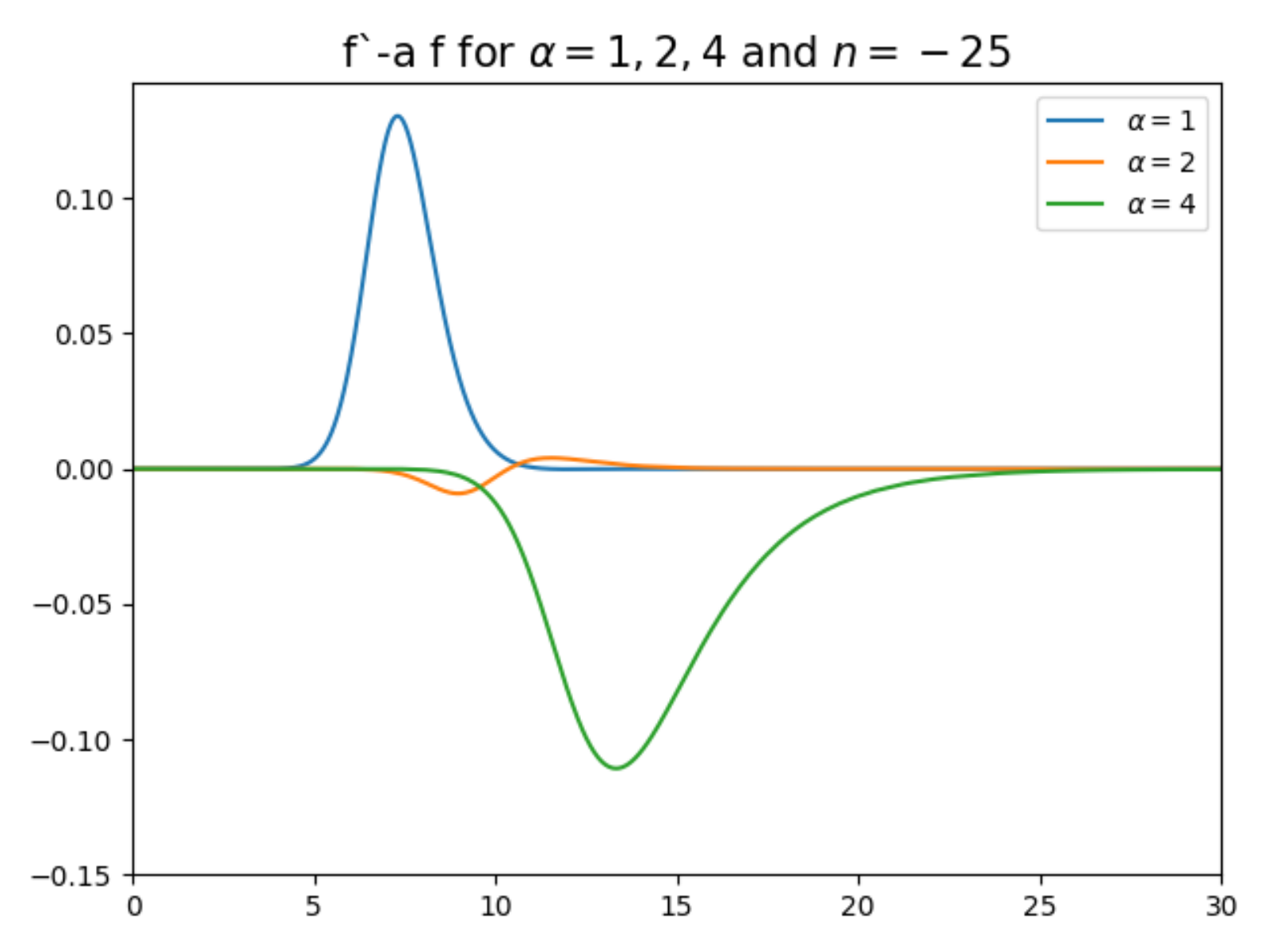}\qquad\includegraphics[width=2.7in]{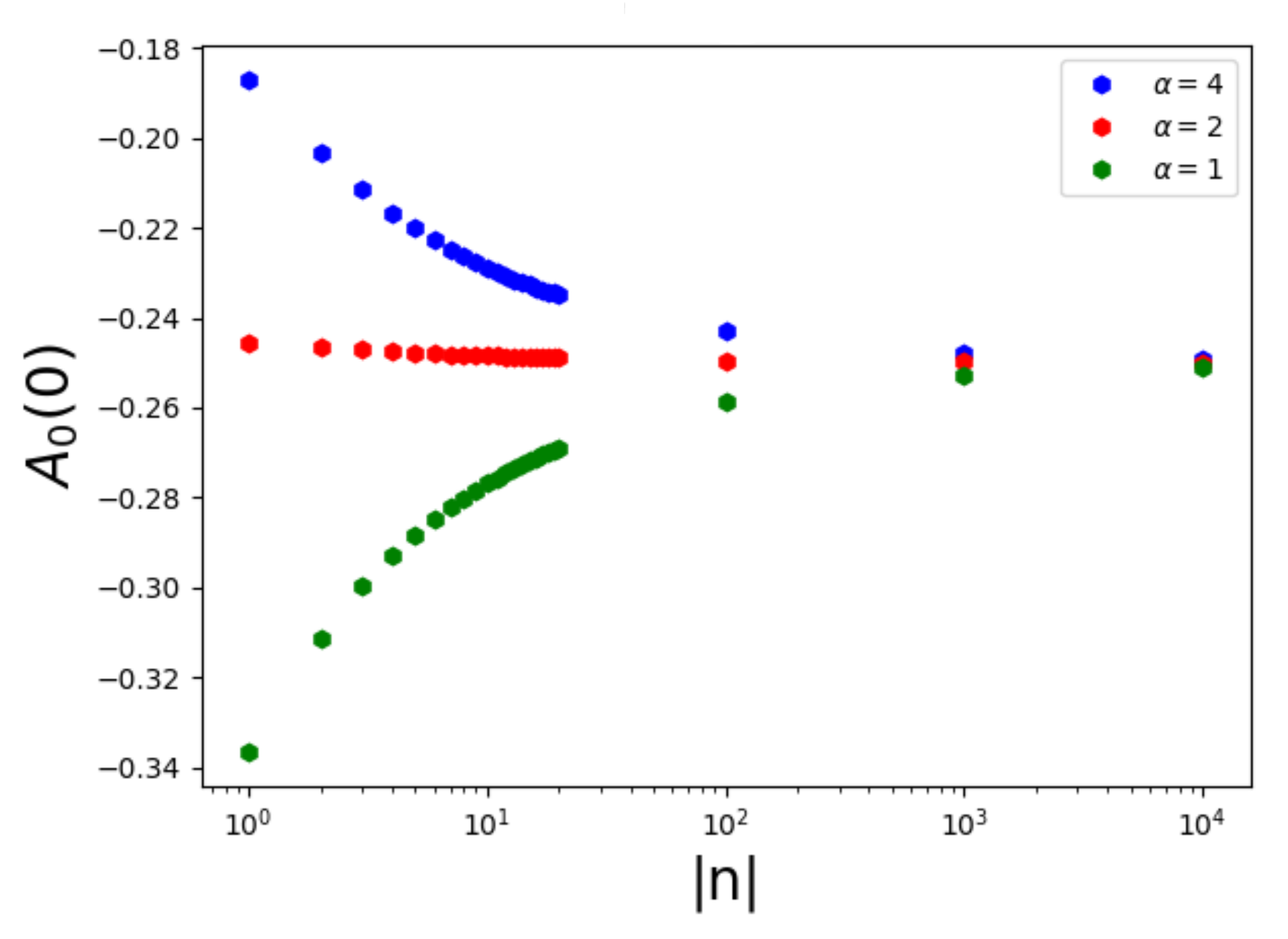}
     \caption{{\small {\it Left: }$\left(\tilde f^\prime - a \tilde f\right)$ plotted for 3 values of $\alpha$ including critical case. $\alpha=2$. {\it Right:} The value of $\tilde A_0$ at the origin for different $\alpha$ as a function of $|n|$. }} \label{fig:error}
    \end{center}
\end{figure}

A surprising feature of the numerical results is how closely the energies of the vortices with $\alpha=2$  match the BPS result $2\pi |n| v^2$ even for low values of $|n|$. This matching is corroborated by  figure \ref{fig:firstorder}, which shows that the vortex profile for $\alpha=2$  almost solves the first order equation $\tilde f' = a \tilde f$. This figure also demonstrates that the vector potential inside the vortex closely follows the result for a uniform magnetic field.
The extent of the departure of the vortex profile from an exact solution to the  first order equation $\tilde f'=a f$ is shown in figure \ref{fig:error}.  Evaluated on the $\alpha=2$ solution, the quantity $(\tilde f^\prime - a \tilde f)$ deviates minimally from zero near the edge of the vortex.

\begin{figure}[h]
\begin{center}
 \includegraphics[width=2.7in]{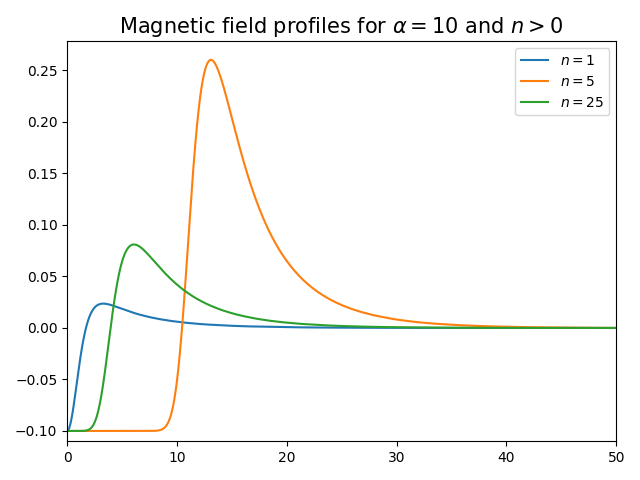} \includegraphics[width=2.7in]{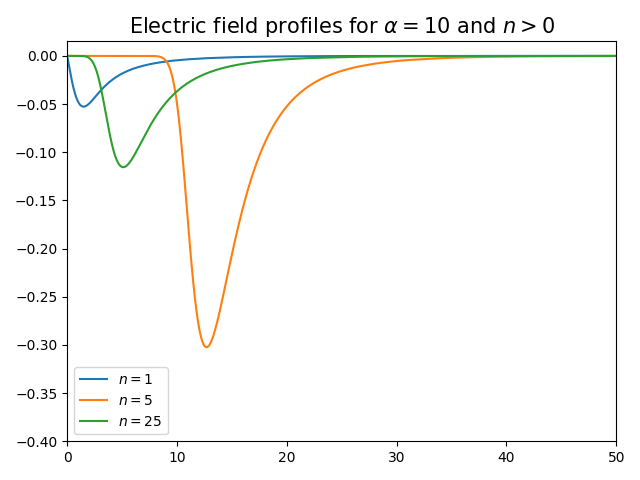}
     \caption{{Magnetic and electric fields for positive flux vortices have support near the edge of the vortex and their peak values grow without bound as $n$ is increased.}} \label{positiveprofile}
    \end{center}
\end{figure}

\begin{figure}[h]
\begin{center}
 \includegraphics[width=2.8in]{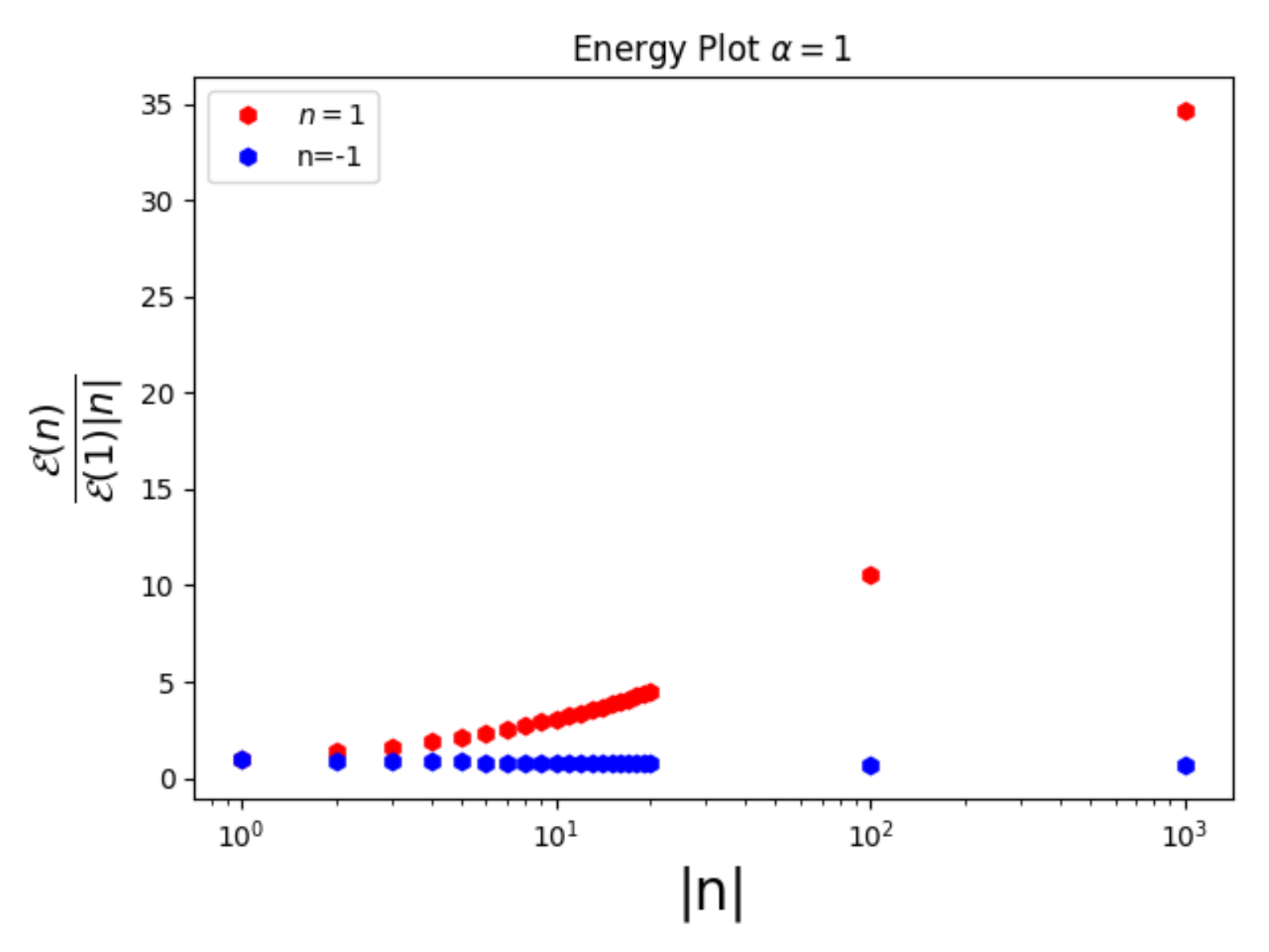}
     \caption{{\small Positive flux vortices have higher energy than their negative flux counterparts.}} \label{positiveenergy}
    \end{center}
\end{figure}

Yet another measure of the relevance of the first order equation $\tilde f^\prime = a \tilde f$ for the $\alpha=2$ solutions is given by the value of $\tilde A_0(0)$. The value of the electrostatic potential at the origin is not fixed as a boundary condition, but an output of the solution. Let us use the equation of motion for the electric field \eqref{eq:dimlesseom2} in conjunction with the first order equation at $\alpha=2$,
\be
\tilde A_0^\prime\,=\,\frac{1}{2}a\tilde f^2 \quad\xrightarrow{\tilde f^\prime = a \tilde f}  \quad 
\tilde A_0\,=\,\frac{1}{4}\left(\tilde f^2-1\right)\,,
\ee
where the integration constant on the right hand side is fixed by requiring that $\tilde A_0$ vanishes as $r\to\infty$. 
We therefore arrive at a prediction,
\be
\tilde A_0(0)\big|_{\alpha=2}\,=\,-\frac{1}{4}\,.
\ee
This is precisely what we see in figure \ref{fig:error} for the $\alpha=2$ solution. However, we also find the unexpected feature that $\tilde A_0(0)$ for other values of $\alpha$ approaches $-1/4$ at large $|n|$. This suggests  that solutions with generic $\alpha$ and large $|n|$ are not approximate  solutions to the first order equation\footnote{For generic $\alpha$, eq. \eqref{eq:dimlesseom2} along with the first order equation $\tilde f^\prime = a \tilde f$, would imply $\tilde A_0(0)=-1/2\alpha$. }.
\paragraph{Positive flux solutions:}
We have explained how positive flux vortex solutions are qualitatively distinct from negative $n$ solutions. In certain limits (large-$\alpha$) they closely resemble  Chern-Simons vortex solutions in vacuum. 
The majority of the flux resides in the ring region or edge of the solution as $n$ is increased (see figure \ref{positiveprofile}).
In figure \ref{positiveenergy} the dependence of the $n$-vortex energy on $|n|$, is displayed for the negative and positive flux solutions, and as expected the latter are more massive.  It is surprising  that ${\cal E }(n)/|n|$ appears to grow faster than $|n|$.
\subsection{Sextic potential ($s=3$)}
\begin{figure}[h]
\begin{center}
 \includegraphics[width=2.9in]{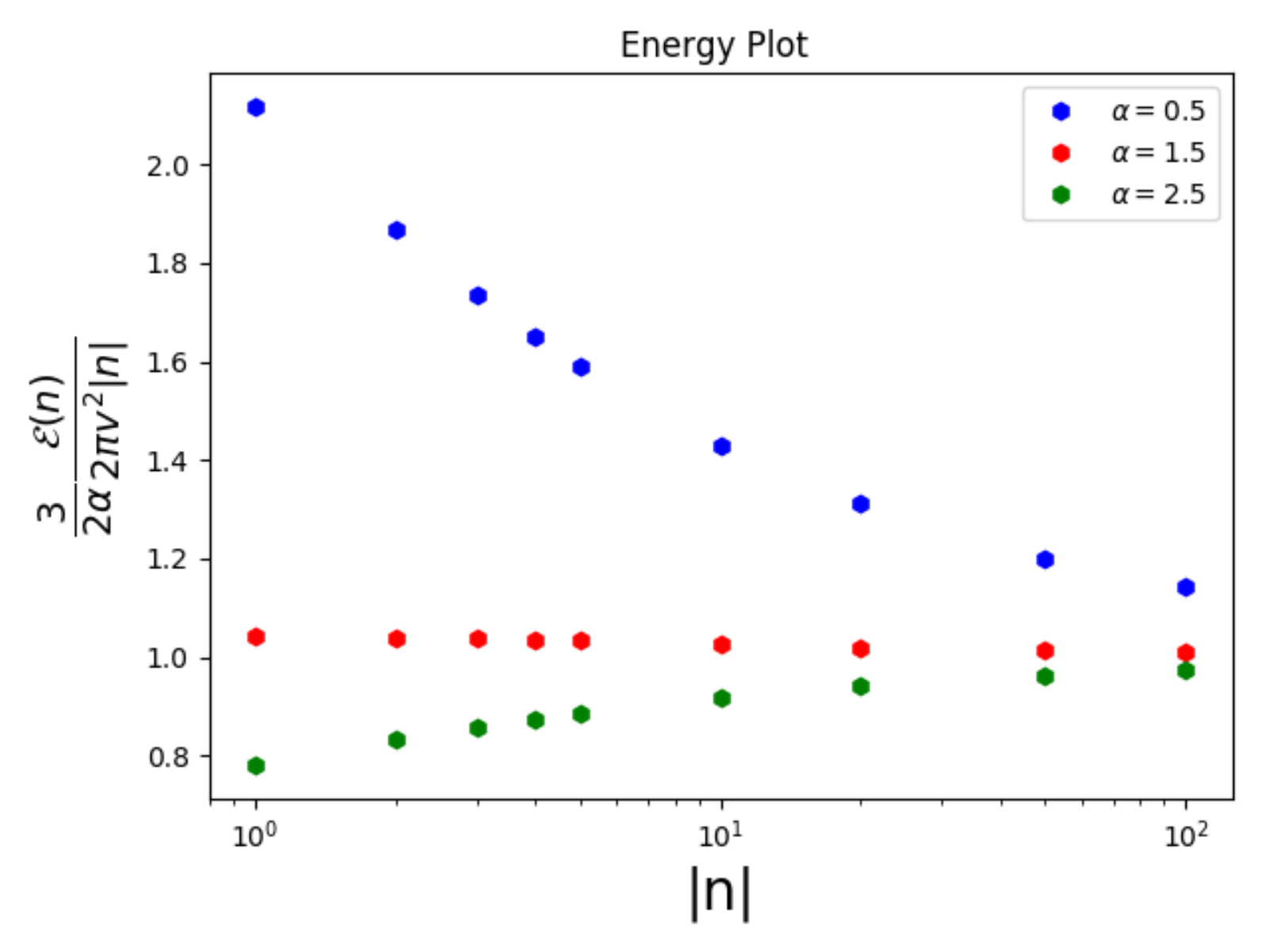}\hspace{0.1in}\includegraphics[width=2.9in]{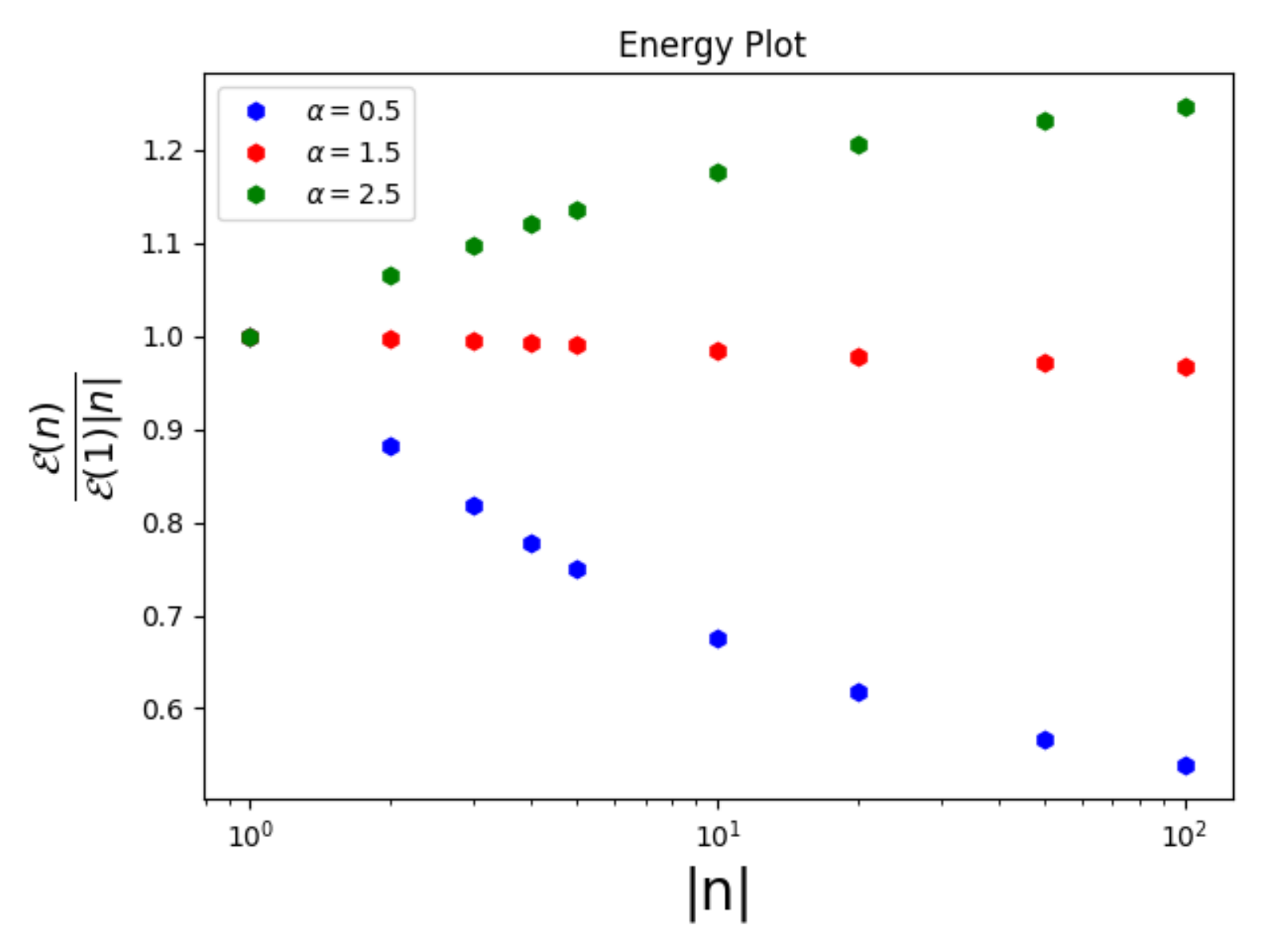}
     \caption{{ \small The scaling of the energy of the $n$-vortex with $|n|$ and $\alpha$, with sextic potential.}} \label{sexticenergy}
    \end{center}
\end{figure}

\begin{figure}[h]
\begin{center}
 \includegraphics[width=2.9in]{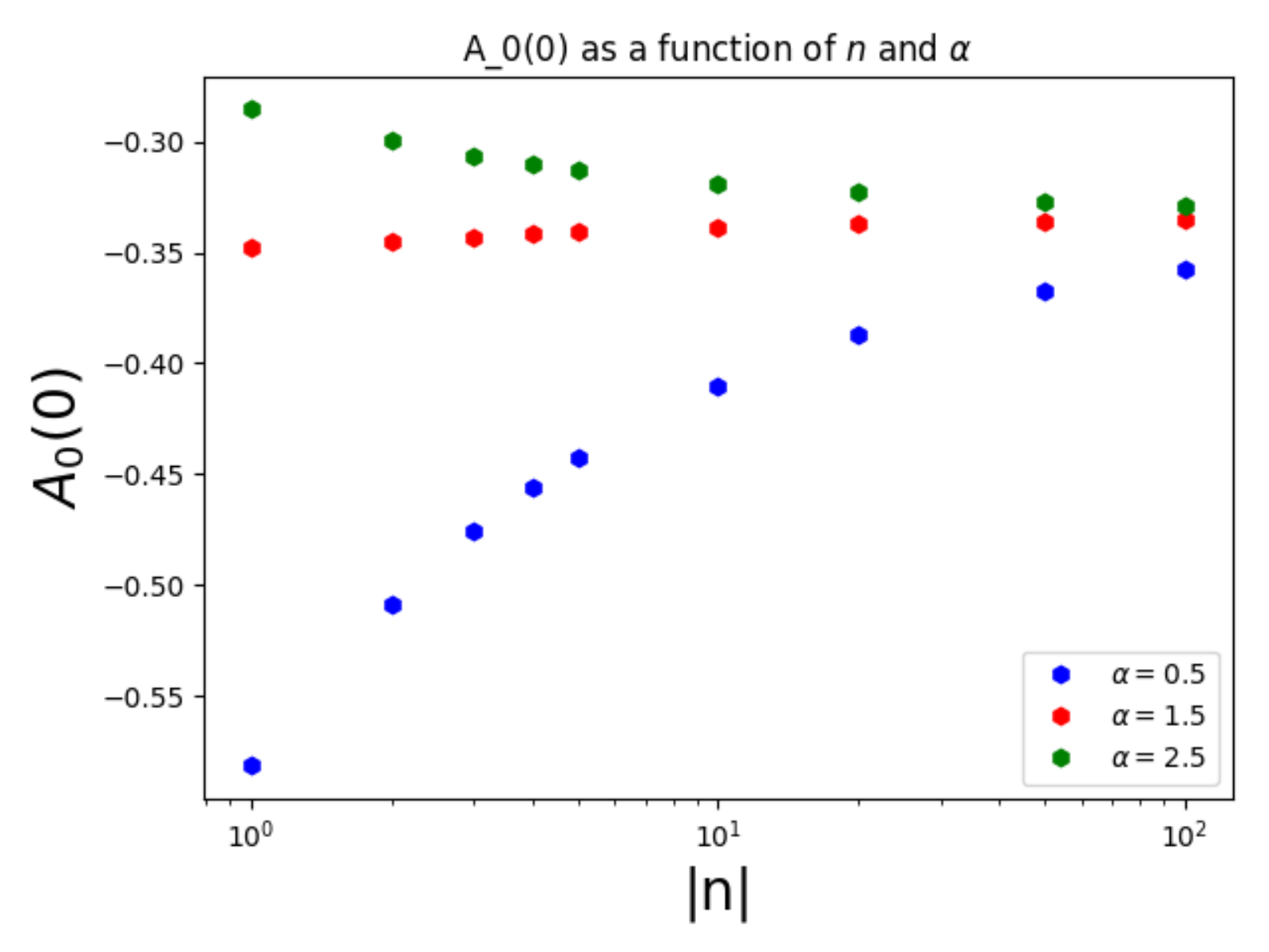}     \caption{{ \small Value of $\tilde A_0$ at the origin for different values of $\alpha$, as a function of $|n|$,  
 with the sextic potential.}} \label{A0sextic}
    \end{center}
\end{figure}

Finally we turn to numerical results for the higher power law potentials, in particular the $s=3$
or sextic potential. We do not expect to see major qualitative differences overall. One special feature of the quartic potential $(s=2)$ is that when $\alpha=2$, there is a precise cancellation of the scalar potential energy contribution to the energy functional. This is not the case for general power laws. Nevertheless there is an approximate cancellation when evaluated on the vortex background at the critical value of $\alpha = \frac{s}{s-1}$.   The critical value for the sextic potential is $\alpha = \frac32$.
\begin{figure}[h]
\begin{center}
 \includegraphics[width=1.95in]{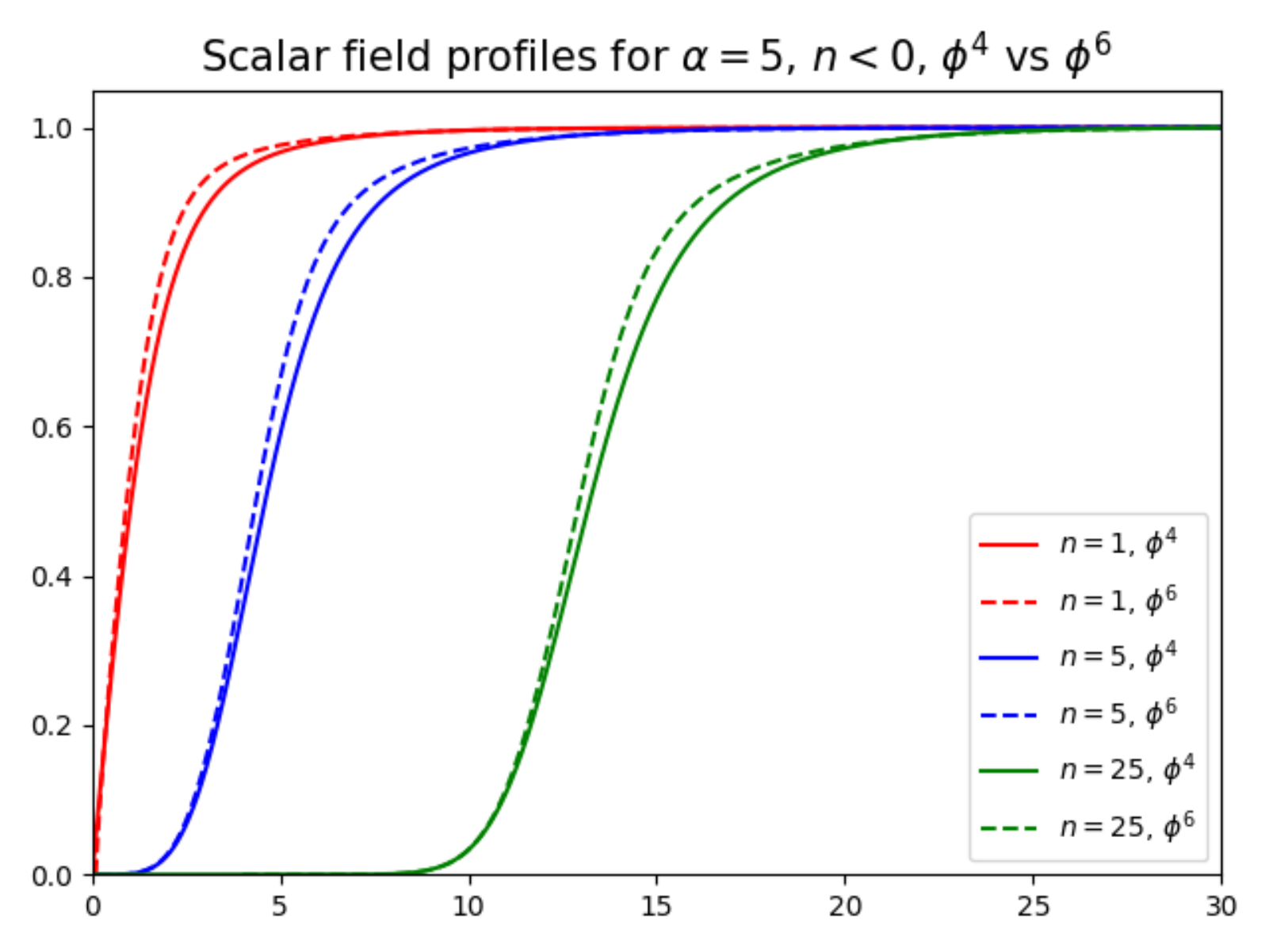}\includegraphics[width=1.95in]{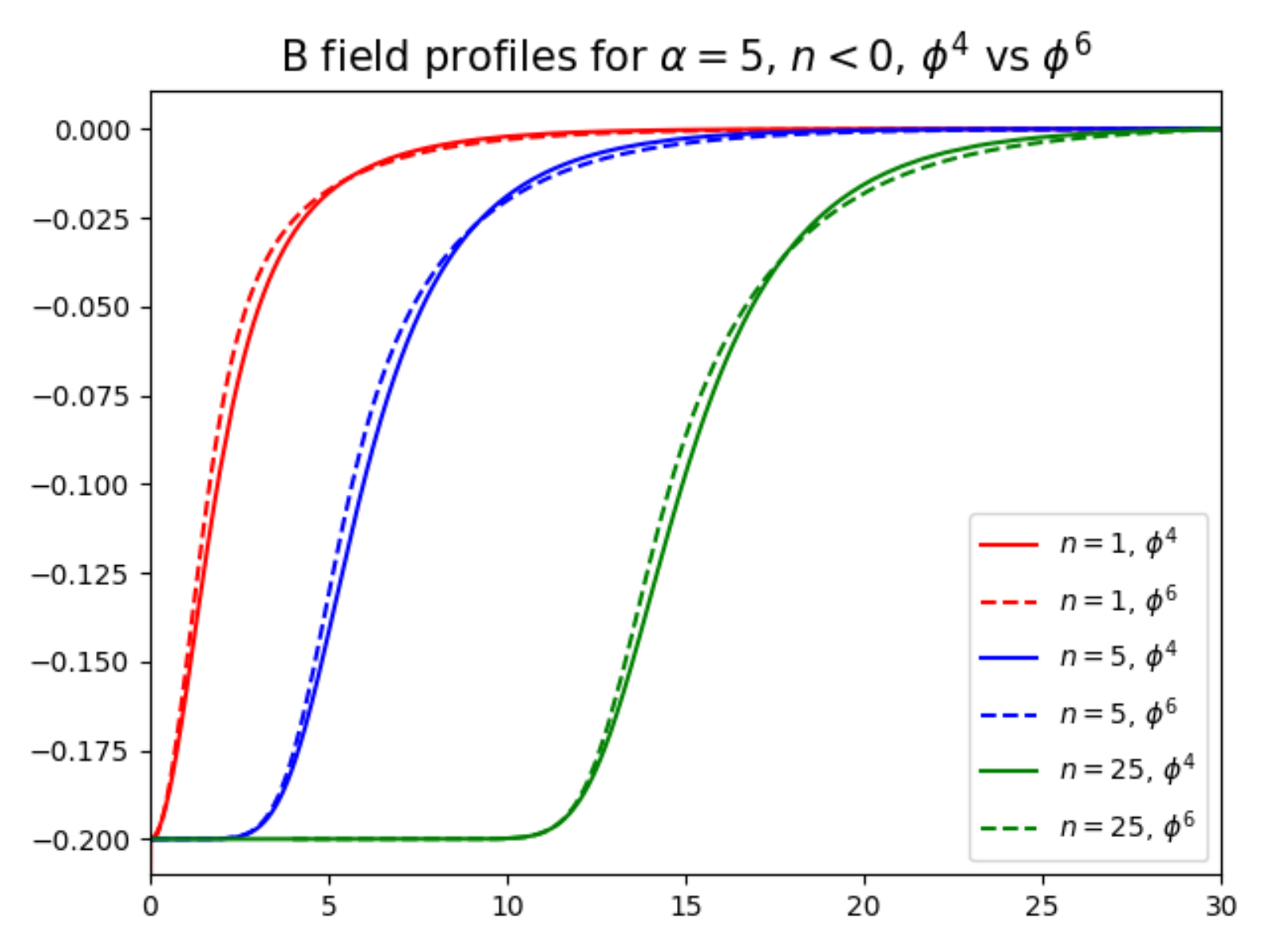}
 \includegraphics[width=1.95in]{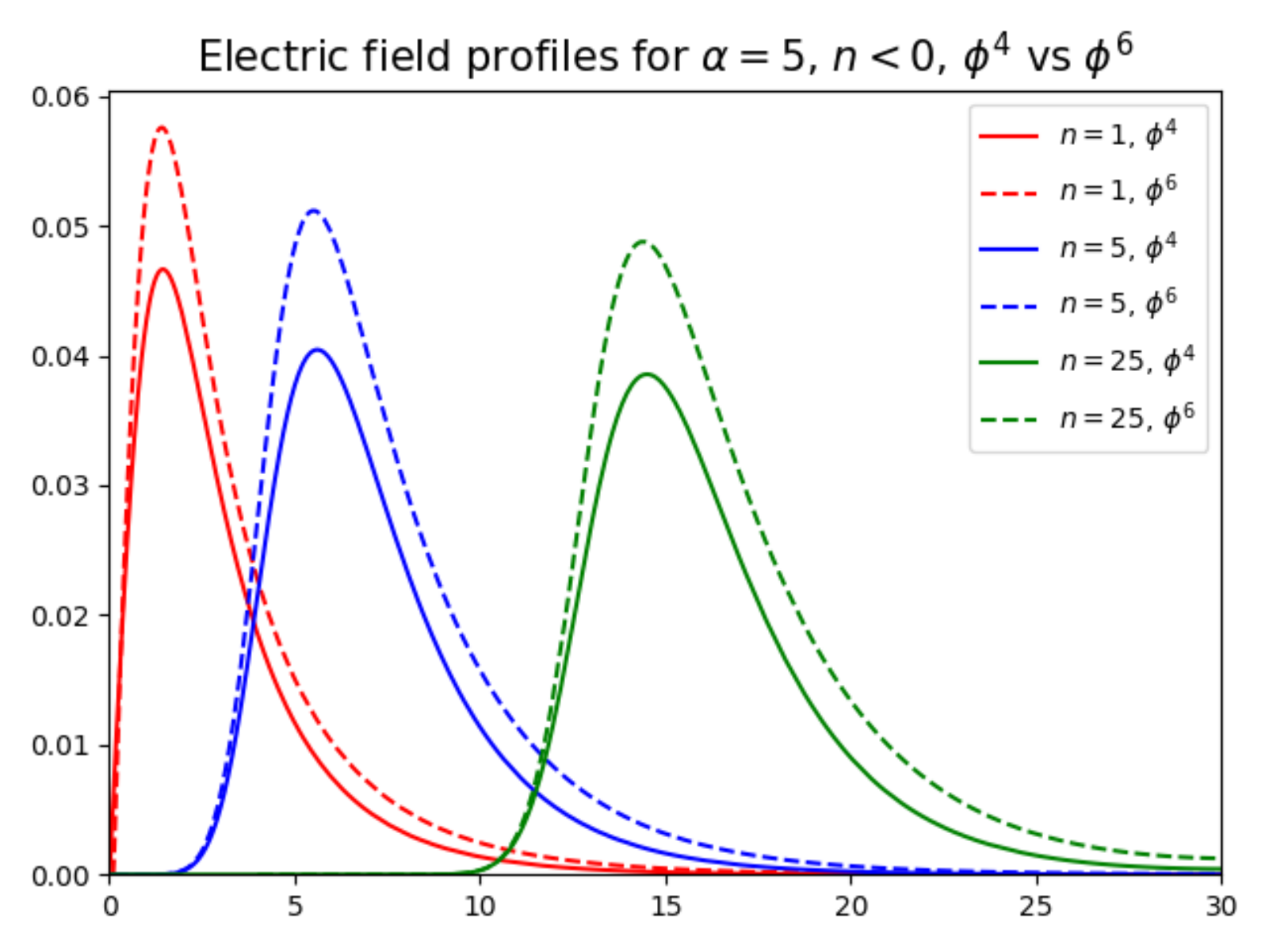}
     \caption{{ \small Vortex profiles with negative winding number for quartic(solid) and sextic(dashed) potentials.}} \label{sexticprofile}
    \end{center}
\end{figure}
The profiles for the negative flux vortex with sextic and quartic potentials are shown on the same plot in figure \ref{sexticprofile}. For the same values of the dimensionless parameter $\alpha$ there is very little difference between the two systems. The transition between the two phases is slightly steeper for the sextic potential. The ratios of the energies of the $n$-vortex to single vortex and and the BPS value ($2\pi v^2 n$) are shown in figure 
\ref{sexticenergy}.
The dependence of the energy on $|n|$ and $\alpha$ matches the prediction \eqref{generalenergy}, and we find a transition line between type I and type II vortices at the critical coupling $\alpha_c = \tfrac32$ which represents the marginally bound ``BPS" case for the sextic potential.   Again the value of $\tilde A_0(0)$ indicates that the profiles almost satisfy the first order equation which would imply,
\be
\tilde A_0(0)\,=\,-\frac{1}{2\alpha_c}\,=\,-\frac13\,.
\ee
This is indeed what we observe in figure \ref{A0sextic}. All large $n$ vortex profiles approach this value
\section{Discussion}
We have studied Abelian Chern-Simons vortices in the presence of a chemical potential driving the theory into the Higgs phase. The numerical solutions reveal several features which are in line with the physical picture presented in an analytically solvable (nonrelativistic) supersymmetric model \cite{Tong:2015xaa, Tong:2003vy}.  The configurations with large (negative) flux show precise BPS-like scaling of energy/grand potential and appear (numerically) to closely solve first order equations.
It would be interesting to make use of the large flux limit to understand the edge excitations of the vortex droplet. We expect that some of the lessons learnt from analyzing this system will be of use in $SU(N)$ and $U(N)$ Chern-Simons-scalar theories with a particle number chemical potential. In these theories the ground state putatively breaks rotational invariance due to condensation of vector fields \cite{Kumar:2018nkf}, and depending on whether we are in the $SU(N)$ or $U(N)$ theory, particle number is ungauged or gauged,  and we will have superfluid or superconducting vortices. 

\acknowledgments We thank Jeff Murugan for discussions on closely related topics. SPK acknowledges support from STFC grant ST/P00055X/1. SS is supported by an STFC studentship under the DTP grant ST/N504464/1.

\end{document}